\documentclass[superscriptaddress,amssymb,amsmath,nobibnotes,aps,prd,eqsecnum,showpacs,
nofootinbib]{revtex4}
\usepackage{amsmath}
\usepackage{amsfonts}
\usepackage{amssymb}
\usepackage{graphicx}

\newcommand{\be}{\begin{eqnarray}}
\newcommand{\ee}{\end{eqnarray}}
\newcommand{\bea}{\begin{eqnarray}}

\newcommand{\eea}{\end{eqnarray}}

\def\be{\begin{equation}}
\def\ee{\end{equation}}
\def\bea{\begin{eqnarray}}
\def\eea{\end{eqnarray}}

\begin{document}

\title{Cosmological models in modified gravity theories with extended nonminimal
derivative
couplings}

\author{Tiberiu Harko}
\email{t.harko@ucl.ac.uk}
\affiliation{Department of Physics, Babes-Bolyai University, Kogalniceanu Street,
Cluj-Napoca 400084, Romania}
\affiliation{Department of Mathematics, University College London, Gower
Street, London
WC1E 6BT, United Kingdom}

\author{Francisco S. N. Lobo}
\email{fslobo@fc.ul.pt}
\affiliation{Instituto de Astrof\'{\i}sica e Ci\^{e}ncias do Espa\c{c}o, Faculdade de
Ci\^encias da Universidade de Lisboa, Edif\'{\i}cio C8, Campo Grande,
P-1749-016 Lisbon, Portugal}

\author{Emmanuel N. Saridakis}
\email{Emmanuel\_Saridakis@baylor.edu}
\affiliation{Instituto de F\'{\i}sica, Pontificia Universidad de Cat\'olica de
Valpara\'{\i}so,
Casilla 4950, Valpara\'{\i}so, Chile}
\affiliation{CASPER, Physics Department, Baylor University, Waco, TX
76798-7310, USA}
\affiliation{Physics Division, National Technical
University of Athens, 15780 Zografou Campus, Athens, Greece}

\author{Minas Tsoukalas}
\email{minasts@central.ntua.gr}
\affiliation{Physics Division, National Technical
University of Athens, 15780 Zografou Campus, Athens, Greece}


\begin{abstract}
We construct gravitational modifications that go beyond Horndeski, namely theories with
extended
nonminimal derivative couplings, in which the coefficient functions depend not only on
the
scalar
field but also on its kinetic energy. Such theories prove to be ghost-free in a
cosmological
background. We investigate the early-time cosmology and show that a de Sitter
inflationary
phase
can be realized as a pure result of the novel gravitational couplings. Additionally, we
study the
late-time evolution, where we obtain an effective dark energy sector which arises from
the
scalar
field and its extended couplings to gravity. We extract various cosmological observables
and
analyse their behavior at small redshifts for three choices of potentials, namely, for
the
exponential, the power-law, and the Higgs potential. We show that the Universe passes
from
deceleration to acceleration in the recent cosmological past, while the effective
dark-energy
equation-of-state parameter tends to the cosmological-constant value at present. Finally,
the
effective dark energy can be phantom-like, although the scalar field is canonical, which
is an
advantage of the model.
\end{abstract}

\maketitle

\section{Introduction}

Horndeski's theory \cite{Horndeski:1974wa} is the most general single-scalar tensor
theory
that has second-order field equations, both for the metric and the scalar field in four
dimensions. It was originally discovered in 1974, then rediscovered independently
\cite{Deffayet:2011gz}, and recently been brought back to attention
\cite{fab4,Charmousis:2011bf,Kobayashi:2011nu} (for a review see
\cite{Charmousis:2014mia}). The generality of the theory is reminiscent of Lovelock's
theorem \cite{Lovelock:1971yv} and it comes as no surprise that many of its terms,
especially those that involve derivative couplings of the scalar with curvature terms,
come from a dimensional reduction of higher dimensional Lovelock theories
\cite{VanAcoleyen:2011mj}. Note that having second-order field equations is crucial, in
order to avoid Ostrogradski instabilities  \cite{ostro,Woodard:2006nt,Woodard:2015zca}.

The advantage of Horndeski cosmological models is that they are able to screen the vacuum
energy coming from any field theory, assuming that after this screening the space should
be in a de Sitter vacuum \cite{Martin-Moruno:2015bda,Martin-Moruno:2015eqa}. These models
allow us to understand the current accelerated expansion of the Universe as the result of
a dynamical evolution towards a de Sitter attractor \cite{Martin-Moruno:2015lha}.
Thus, it was shown that Horndeski models with a de Sitter critical point for any kind of
material content may provide a mechanism to alleviate the cosmological constant problem
\cite{Martin-Moruno:2015kaa}.  The cosmological scenario that results when
considering the radiation and matter content was also studied, and it was concluded that
their background dynamics is compatible with the latest observational data.

Despite the huge interest in these theories, extensions of Horndeski's theory have also
been recently discussed. In \cite{Gleyzes:2014dya} a new class of scalar-tensor theories
was introduced, going beyond Horndeski's theory, where despite the fact that the
equations
of motion contain higher derivatives, they can be cast in a way that they contain only
second-order ones \cite{Zumalacarregui:2013pma}. Additionally, these generalized theories
were shown to be free of ghost instabilities in the unitary gauge \cite{Gleyzes:2014qga},
and later on this was also verified using the Hamiltonian formalism
\cite{Domenech:2015tca,Langlois:2015cwa,Deffayet:2015qwa,Langlois:2015skt,
Crisostomi:2016tcp}, due to  the existence of a primary constraint
which prevents the propagation of extra degrees of freedom \cite{Crisostomi:2016tcp} (see
also \cite{Gao:2014soa} and \cite{Crisostomi:2016czh,Ezquiaga:2016nqo,BenAchour:2016fzp}
for
additional
descriptions).
We mention that  these extended theories can also address the cosmological constant
problem \cite{Babichev:2015qma} via a self-tunning mechanism, similarly to the analysis
done in the original Horndeski theory for the so-called \textit{Fab Four} theory
\cite{fab4,Charmousis:2011bf} (the cosmological aspects of the Fab-Four have been
explored
in \cite{Copeland:2012qf}). A detailed analysis of the cosmological self-tunning
and local solutions in the context of beyond Horndeski theories has also been
explored in \cite{Babichev:2016kdt}. Recently, it was shown that the two additional
Lagrangian pieces, appearing in theories
beyond Horndeski, could be re-expressed in a very elegant
and compact way, by allowing the potentials to also depend on the kinetic term of the
scalar field \cite{Babichev:2015qma}.

One interesting subclass of Horndeski theory, which has been given much attention
recently, includes the nonminimal (kinetic) coupling of matter to gravity by inserting
derivative couplings between the geometry and the kinetic part of the scalar
field \cite{Amendola:1993uh}, which leads to interesting new dynamical cosmological
phenomena \cite{Sushkov:2009hk,Saridakis:2010mf}, including the existence of an effective
cosmological constant \cite{Capozziello:1999uwa,Capozziello:1999xt}.
The nonminimal derivative coupling leads to cosmological models with rich phenomenology,
such as solutions containing a Big Bang, expanding Universes with no beginning,
cosmological bounces, eternally contracting Universes, a Big Crunch, and a Big Rip
avoidance \cite{Saridakis:2010mf,Granda:2010ex,Granda:2010hb,Sadjadi:2010bz,
Sami:2012uh,Banijamali:2012kq,Bruneton:2012zk,Sheikhahmadi:2016wyz}. In particular, it
was
shown that one is able to explain in a unique manner both a quasi-de Sitter phase and an
exit from it without any fine-tuned potential \cite{Sushkov:2009hk}. Furthermore, one can
successfully describe the sequence of cosmological epochs  without any fine-tuned
potential \cite{Sushkov:2012za}.
Using couplings of this type, it was found that in the absence of other matter sources
or in the presence of only pressureless matter, the scalar field behaves as pressureless
matter and its sound speed is vanishing \cite{Gao:2010vr}. These properties
enable the scalar field to be a candidate of cold dark matter. It was also shown that if
the kinetic term is coupled to more than one Einstein tensor, then the equation of
state is always approximately equal to $-1$, independently from the potential flatness,
and hence the scalar may also be considered a candidate for the inflaton.
Tachyon models involving nonminimal derivative coupling have also been explored
\cite{Shchigolev:2011nma,Banijamali:2011qb}, while Chaplygin gas model in this framework
were studied in \cite{Granda:2011zy}. Moreover, the dynamics of entropy perturbations in
the two-field assisted dark energy model with mixed kinetic terms was also studied  in
\cite{Karwan:2010xw}. Recently there has also been an investigation on how the
derivative
coupling can mimic cold dark matter at cosmological level and also explain the flattening
of
galactic rotation curves \cite{Rinaldi:2016oqp}.

The inflationary context within this theory has been extensively analysed too. In the
case of a power-law potential, and using the dynamical system method,
all possible asymptotical regimes of the model were analysed \cite{Skugoreva:2013ooa}. It
was shown that for sloping potentials there exists a quasi-de Sitter asymptotic
corresponding to an early inflationary Universe. In contrast to standard inflationary
scenario, the kinetic-coupling inflation does not depend on a scalar field potential and
is only determined by the coupling parameter.
In addition to this, there is a unique nonminimal derivative coupling of the Standard
Model Higgs boson to gravity which propagates no more degrees of freedom than
General Relativity sourced by a scalar field, and reproduces a successful inflating
background within the Standard Model Higgs parameters and, finally, does not suffer from
dangerous quantum corrections \cite{Germani:2010gm}. The slow-roll conditions have been
found \cite{Granda:2011zk}, and the reheating temperature was obtained
\cite{Sadjadi:2012zp,Sadjadi:2013na} (see also recent analyses in
\cite{Gumjudpai:2016ioy} and \cite{Dalianis:2016wpu}).
Furthermore, the cosmological perturbations originated at the inflationary stage were
studied and the consistency of the results with observational constraints coming from
Planck 2013 data were investigated \cite{Sadjadi:2013psa}.
Moreover, these scenarios exhibit a gravitationally enhanced friction during inflation,
where even steep potentials with theoretically natural model parameters can drive cosmic
acceleration \cite{Tsujikawa:2012mk}, while being compatible with the current
observational data mainly due to the suppressed tensor-to-scalar ratio. Finally, the
gravitational production of heavy $X$-particles of mass of the order of the inflaton
mass, produced after the end of inflation, was also studied \cite{Koutsoumbas:2013boa},
where it  was found that  this production is suppressed as the strength of the coupling
is increased.

A combined perturbation and observational investigation of the scenario of nonminimal
derivative coupling between a scalar field and curvature was performed in
\cite{Dent:2013awa}. Using Type Ia Supernovae (SNIa), Baryon Acoustic Oscillations (BAO),
and Cosmic Microwave Background (CMB) observations, it was shown that, contrary to its
significant effects on inflation, the nonminimal derivative coupling term has a
negligible
effect on the Universe acceleration, since it is driven solely by the usual scalar-field
potential. Therefore, the scenario can provide a unified picture of early and late time
cosmology, with the nonminimal derivative coupling term responsible for inflation, and
the
usual potential responsible for late-time acceleration.

Finally, nonminimal derivative couplings to gravity have also been explored in a variety
of extended theories of gravity. For instance, one can incorporate an additional
coupling to the Gauss Bonnet invariant, obtaining rich cosmological behavior, with both
decelerated and accelerated phases \cite{Granda:2011eh,Granda:2012hm}. Additionally, a
large class of scalar-tensor models with interactions containing the second derivatives
of
the scalar field but not leading to additional degrees of freedom have also been
extensively investigated \cite{Deffayet:2010qz}. These models exhibit peculiar features,
such as an essential mixing of scalar and tensor kinetic terms, named kinetic braiding,
and possess a rich cosmological phenomenology, including a late-time asymptotic de Sitter
state, and a possible phantom-divide crossing, with neither ghosts nor gradient
instabilities. Finally, the nonminimal derivative coupling to gravity has also been
investigated in the context of the curvaton model \cite{Feng:2013pba}, or in the
framework of $N = 1$ four-dimensional new-minimal supergravity \cite{Farakos:2012je}.

In this work we are interested in investigating a theory that goes beyond Horndeski,
based
on a generalization of nonminimal derivative coupling. In particular, we consider the
latter coupling and introduce an additional arbitrary coefficient-function of the field
and its derivatives. We mention that this class is not included in Horndeski
theory, since only specific combinations of it are allowed
\cite{Deffayet:2011gz,Gleyzes:2013ooa,Gleyzes:2014rba}.
This paper is outlined in the following manner. In Section \ref{sec1}, we present the
action and deduce the gravitational field equations. In Section \ref{sec2}, we apply the
developed formalism to a spatially flat Friedmann-Robertson-Walker (FRW) background
metric, and present the modified Friedmann equations. The early-time cosmology is briefly
analysed in Section \ref{sec3}, and the late-time evolution is considered in Section
\ref{sec4}. In the latter, we study the full Friedmann equations  and focus on important
observables, by considering three well-known scalar potentials, such as the exponential,
power-law and the Higgs potential. Finally, in Section \ref{sec5} we discuss our
results and conclude.

\section{Extended nonminimal derivative coupling}\label{sec1}

In this work, we consider a generalized nonminimal coupling of the scalar field
derivative
to gravity, by introducing an additional arbitrary coefficient-function of the field and
its derivatives. The action is given by
\begin{eqnarray}  \label{action}
S=\int d^4x\sqrt{-g}\left\{ \frac{R}{16\pi G} -\frac{1}{2}\Big\{
g_{\mu\nu}-\big[\beta + \zeta F(X,\phi)\big]G_{\mu\nu} \Big\}\nabla^{\mu}\phi \,
\nabla^{\nu}\phi -
V(\phi)\right\} +S_m,
\end{eqnarray}
with $g_{\mu\nu}$ the metric, $g=\det(g_{\mu\nu})$, $R$ the scalar
curvature, $\beta$ and $\zeta$ the derivative coupling parameters, $V(\phi)$ the
scalar field potential, $X=-\frac{1}{2} \nabla^{\mu} \phi \,  \nabla_{\mu}\phi$ the
scalar
kinetic energy and $F(X,\phi)$ an arbitrary function. Note that in principle $\beta$
can be absorbed inside $ \zeta F(X,\phi)$, however we prefer to keep it separately in
order to be able to reproduce at any stage the results of the simple nonminimal
derivative coupling by setting $\zeta$ to $0$.
Finally, we have included the
usual matter action, corresponding to a matter fluid of energy density $%
\rho_m$ and pressure $p_m$.

Variation of the action with respect to the metric leads to the field
equations
\begin{eqnarray}  \label{eoms1}
\frac{1}{16\pi G}G_{\mu\nu}
-\frac{1}{2} \, \nabla_{\mu}\phi  \, \nabla_{\nu}\phi +\frac{1}{4}%
g_{\mu\nu}  \nabla^{\alpha}\phi  \, \nabla_{\alpha}\phi
+\frac{\beta}{2}
\Bigg\{-\frac{1}{2}%
g_{\mu\nu}G^{\alpha\beta}\nabla_{\alpha}\phi\nabla_{\beta}\phi+2G_{(\mu}^{\,
\,\,\,\,\lambda}\nabla_{\nu)}\phi\nabla_{\lambda}\phi+\frac{1}{2}%
R\nabla_{\mu}\phi\nabla_{\nu} \phi
	\nonumber \\
-\frac{1}{2} R_{\mu\nu}\nabla^{\alpha}\phi%
\nabla_{\alpha}\phi
 +\frac{1}{2}g_{\mu\nu}
\left[(\square\phi)^{2}-\nabla_{\alpha}%
\nabla_{\beta}\phi\nabla^{\alpha} \nabla^{\beta}
\phi-R_{\alpha\beta}\nabla^{\alpha}\phi\nabla^{\beta}\phi\right]
+\nabla_{%
\mu}\nabla^{\alpha}\phi\nabla_{\nu}
\nabla_{\alpha}\phi-\square\phi\nabla_{\mu}\nabla_{\nu}\phi
	\nonumber \\
+R_{\mu\,\,\,%
\nu}^{\,\,\,\alpha\,\,\,\,\beta} \nabla_{\alpha} \phi\nabla_{ \beta}\phi %
\Bigg\}
+\frac{\zeta}{2}
\left[\frac{1}{2}\left(XFG_{\mu\nu}+ 2 F
P_{\mu\,\,\,\nu}^{\,\,\,\alpha\,\,\,\beta}\nabla_{\alpha}\phi  \, \nabla_{\beta}\phi
\right)
+ F_{,X}  \nabla_{\mu} \phi  \, \nabla_{\nu}\phi \;
G^{\alpha\beta} \; \nabla_{\alpha} \phi  \, \nabla_{\beta}\phi \right]
	\nonumber \\
- \frac{1}{2}\epsilon_{\mu} ^{\,\,\,\alpha\sigma\gamma}
\epsilon_{\nu\,\,\,\,\,\,\gamma}^{\,\,\,\beta\rho}
\left[ F\,  \nabla_{\beta} \nabla_{\alpha}\phi \; \nabla_{\sigma} \nabla_{\rho}\phi +2
\nabla_{(\alpha}F \, \nabla_{\beta)}\phi  \, \nabla_{\sigma} \nabla_{\rho}\phi -
\nabla_{\alpha}\phi  \, \nabla_{\beta}\phi    \nabla_{\sigma} \nabla_{\rho}F \right]
+\frac{%
1}{2}V( \phi)g_{\mu\nu} = -\frac{1}{\sqrt{-g}}\frac{\delta S_m}{\delta
g^{\mu\nu}} \,,
\end{eqnarray}
where $P_{\mu\nu\alpha \beta}$ is the double dual of the Riemann tensor defined as
\cite{mtw}
\begin{eqnarray}
P^{\mu\nu}_{\,\,\,\,\,\,\,\,\alpha\beta}\equiv\frac{1}{4}\delta^{\mu\nu%
\gamma\delta}_{\sigma\lambda\alpha\beta}R^{\sigma\lambda}_{\,\,\,
\,\,\,\gamma\delta}=R^{\mu\nu}_{\,\,\,\,\,\,\alpha\beta}-2R^{\mu}_{\,\,\,[%
\alpha}\delta^{\nu}_{\,\,\beta]}+2R^{\nu}_{\,\, \,[\alpha}\delta^
{\mu}_{\,\, \beta]}+R\delta^{\mu}_{\,[\alpha}\delta^{\nu}_{\,\beta]},
\end{eqnarray}
and where $\nabla_{(\mu}\phi R^{\alpha}_{\nu)} = \frac{1}{2}(\nabla_{\mu}\phi
R^{\alpha}_{\nu}+\nabla_{\nu}\phi R^{\alpha}_{\mu}) $. Additionally,
variation of the action (\ref{action}) with respect to $\phi$ provides the
scalar field equation of motion, namely
\begin{eqnarray}
\label{eomphi1}
\square\phi-\beta G^{\mu\nu}\nabla_{\mu}\nabla_{\nu}\phi
-\frac{\zeta}{2}\Big[%
2\nabla_\mu\left( F G^{\mu\nu}\nabla_{\nu}\phi \right) + 2 \nabla_\alpha\left(
F_{,X} G^{\mu\nu}  \nabla_{\mu} \phi \, \nabla_{\nu}\phi \,\nabla^{\alpha}\phi  \right) -
F_{,\phi}
G^{\mu\nu} \nabla_{\mu} \phi \, \nabla_{\nu}\phi \Big]-V_{\phi}=0.
\end{eqnarray}
In this work we will consider the case where $F(\phi,X)=X$, which is adequate to capture
the new features of the theories beyond Horndeski at hand. Hence the coupling
constant $\zeta$ has dimensions of inverse mass to the power of six while $\beta$ has
dimensions of inverse mass to the power of two.
Using the identity
\begin{eqnarray}
\epsilon_{\mu\alpha\beta\gamma}\epsilon^{\nu\kappa\lambda\rho}=-\delta_{\mu%
\alpha\beta\gamma}^{\nu\kappa\lambda\rho}
\end{eqnarray}
and the fact that apart from the potential term the rest of the action is
shift symmetric, we can now write our field equations in the following
elegant way
\begin{eqnarray}
\label{eoms3}
\frac{1}{16\pi G}G_{\mu}{}^{\nu}-\frac{1}{2} \nabla_{\mu} \phi \, \nabla^{\nu}\phi
+ \frac{1}{4}\delta_{\mu}{}^{\nu} \nabla^{\alpha} \phi \, \nabla_{\alpha}\phi
+\frac{\beta}{2}
\Bigg\{-\frac{1}{2}\delta_{\mu}^{\,\,\,\nu}G^{\alpha\beta}\nabla_{\alpha}
\phi\nabla_{\beta}\phi+2G_{(\mu}^{\,\,\,\,\,\lambda}\nabla^{\nu)}\phi%
\nabla_{\lambda}\phi+\frac{1}{2} R\nabla_{\mu} \phi\nabla^{\nu}\phi
	\nonumber \\
-\frac{1}{%
2}R_{\mu}^{\,\,\,\,\nu}\nabla^{\alpha}\phi\nabla_{\alpha}\phi
+\frac{1}{2}\delta_{\mu}^{\,\,\,\nu}
\left[(\square\phi)^{2}-\nabla_{%
\alpha}\nabla_{\beta}
\phi\nabla^{\alpha}\nabla^{\beta}\phi-R_{\alpha\beta}\nabla^{\alpha}\phi%
\nabla^{\beta}\phi\right]
+\nabla_{\mu} \nabla^{\alpha}
\phi\nabla^{\nu}\nabla_{\alpha}\phi
		\nonumber \\
-\square\phi\nabla_{\mu}\nabla^{\nu}%
\phi+R_{\mu}^{\,\,\,\alpha\nu\beta} \nabla_{\alpha} \phi\nabla_{\beta} \phi %
\Bigg\}
+\frac{\zeta}{2}
\left[\frac{1}{2}\left(X^{2}G_{\mu}^{\,\,\,\nu}+ 2 X
P_{\mu\alpha}{}^{\nu\beta} \nabla^{\alpha} \phi \, \nabla_{\beta}\phi \right)
+ \nabla_{\mu} \phi \, \nabla^{\nu}\phi  G^{\alpha\beta} \nabla_{\alpha} \phi \,
\nabla_{\beta}\phi
\right]
	\nonumber  \\
+\frac{1}{2}\delta_{\mu\alpha\sigma\gamma}^{\nu\beta\rho\gamma}
\left[ X \nabla^{\beta} \, \nabla_{\alpha}\phi
\nabla^{\sigma} \, \nabla_{\rho}\phi
+2\nabla^{(\alpha} X  \nabla_{\beta)}\phi
\nabla^{\sigma}  \, \nabla_{\rho}\phi
- \nabla^{\alpha} \phi \, \nabla_{\beta}\phi   \nabla^{\sigma} \, \nabla_{\rho}X\right]
+\frac{1}{2}V(\phi)\delta_{\mu}^{\,\,\,\nu} =\frac{1}{2}
T_{\mu}^{\,\,\,\nu} \,,
\end{eqnarray}
and
\begin{eqnarray}
\label{eomphi3}
\nabla_{\alpha}J^{\alpha}=V_{\phi},
\end{eqnarray}
where
\begin{eqnarray}
J^{\alpha}=\Big[%
g^{\alpha\beta}-\beta G^{\alpha\beta}-\zeta (X\,G^{\alpha\beta}
+g^{\alpha\beta}\,G^{\mu\nu}\nabla_{\mu}\phi\nabla_{\nu}\phi)\Big]%
\nabla_{\beta}\phi.
\end{eqnarray}

\section{Cosmological Equations}\label{sec2}

In this section we are interested in investigating the cosmological implications of
theories with extended nonminimal derivative couplings. Hence, we focus on a
spatially-flat
Friedmann-Robertson-Walker (FRW) background metric of the form
\begin{equation}
ds^{2}=-dt^{2}+a^{2}(t)\,\delta _{ij}dx^{i}dx^{j},  \label{FRW0metric0}
\end{equation}%
where $t$ is the cosmic time, $x^{i}$ are the comoving spatial coordinates, $%
a(t)$ is the scale factor and $H=\dot{a}/a$ is
the Hubble parameter, (a dot denotes differentiation with respect to $t$).
Additionally, we consider the scalar field to be homogeneous, that is $\phi
=\phi (t)$. Finally, as usual, we consider the matter sector to correspond to a perfect
fluid.

In this case,  the field equations (\ref{eoms3}) provide the two Friedmann equations:
\begin{eqnarray}
H^{2}=\frac{4\pi \,G}{3\left( 1-12\pi G\beta \,\dot{\phi}^{2}+20\pi
\,G\,\zeta \,\dot{\phi}^{4}\right) }\left[ 2V(\phi )+2\rho _{m}+\dot{\phi}%
^{2}\right] \,,   \label{FR1}
\end{eqnarray}%
\begin{equation}
2\dot{H}+3H^{2}+4\pi \,G\left( 2p_{m}-2V+\dot{\phi}^{2}\right) -4\pi
\,G\,\beta \dot{\phi}\Big[(3H^{2}+2\dot{H})\dot{\phi}+4H\ddot{\phi}\Big]%
+4\pi \,G\,\zeta \,\dot{\phi}^{3}\Big[(2\dot{H}+3H^{2})\dot{\phi}+8H\,\ddot{%
\phi}\Big]=0 \,, \label{FR2}
\end{equation}%
while equation (\ref{eomphi3}) gives
\begin{equation}
\ddot{\phi}+3H\dot{\phi}+V_{\phi }+3\,\beta \,H\Big[(3H^{2}+2\dot{H})\dot{%
\phi}+H\ddot{\phi}\Big]-6\,\zeta \,H\dot{\phi}^{2}\Big[(2\dot{H}+3H^{2})\dot{%
\phi}+3H\,\ddot{\phi}\Big]=0.\   \label{FR3}
\end{equation}%
As we can easily see, despite the appearance of higher derivatives in Eq. (\ref{eoms1}),
on
the cosmological background they disappear \cite{Gleyzes:2014dya}, and we only have to
deal with up to two derivatives.
From the above expressions one can see that the Friedmann equations
(\ref{FR1}) and (\ref{FR2}) can be written in the usual form, namely
\begin{eqnarray}
H^{2}&=&\frac{8\pi G}{3}(\rho _{DE}+\rho _{m}) , \label{FR1b} \\
2\dot{H}+3H^{2}&=&-8\pi G(p_{DE}+p_{m}),  \label{FR2b}
\end{eqnarray}%
where we have defined an effective dark energy sector with energy density and pressure:
\begin{equation}
\rho _{DE}\equiv \rho _{\phi }=\frac{\dot{\phi}^{2}}{2}+V(\phi )+H^{2}\dot{%
\phi}^{2}\left( \frac{9}{2}\beta -\frac{15}{2}\zeta \,\dot{\phi}^{2}\right),
\label{rhoDE}
\end{equation}%
\begin{equation}
p_{DE}\equiv p_{\phi }=\frac{\dot{\phi}^{2}}{2}-V(\phi )-\beta \frac{\dot{%
\phi}}{2}\Big[(3H^{2}+2\dot{H})\dot{\phi}+4H\ddot{\phi}\Big]+\zeta \frac{%
\dot{\phi}^{3}}{2}\Big[(2\dot{H}+3H^{2})\dot{\phi}+8H\,\ddot{\phi}\Big],
\label{pDE}
\end{equation}
respectively. Therefore, in the scenario at hand, the dark-energy
equation-of-state parameter is given by
\begin{equation}
w_{DE}\equiv \frac{p_{DE}}{\rho _{DE}}.  \label{wDE}
\end{equation}
One can straightforwardly see that, in terms of the dark energy density and
pressure, the scalar field evolution equation (\ref{FR3}) can be
written in the standard form
\begin{equation}
\dot{\rho}_{DE}+3H(\rho _{DE}+p_{DE})=0.
\end{equation}%
Furthermore, the matter energy density and pressure satisfy the standard
evolution equation
\begin{equation}
\dot{\rho}_{m}+3H(\rho _{m}+p_{m})=0.  \label{rhoevol}
\end{equation}

Finally, we introduce the deceleration parameter $q$, which is an indicator of the
accelerated expansion, and is defined as
\begin{equation}
q=\frac{d}{dt}\frac{1}{H}-1,
\label{deccparam}
\end{equation}
and thus negative values of $q$ correspond to  accelerating evolution. Additionally, in
order to allow for an easy comparison between the observational and
theoretical results, instead of the time variable $t$ we can use the redshift $z$,
defined as
\begin{equation}
1+z=\frac{1}{a},
\label{redshoftfedin}
\end{equation}%
where we have normalized the scale factor $a\left( z\right) $ so that its current value
is $a(0)=1 $. Thus,   time derivatives  can be expressed as
\begin{equation}
\frac{d}{dt}= -H(z)(1+z)\frac{d}{dz}.
\label{timeredshiftrel}
\end{equation}

\section{Early-time cosmology}\label{sec3}

In order to examine the early-time behavior of the scenario at hand we will neglect the
matter content of the theory. We are interested in exponential cosmological
solutions, which could describe the inflationary epoch. Note that since we
desire to study the pure effects of the novel, extended nonminimal
derivative couplings, we do not consider an explicit
potential, since it is well known that a potential term can easily drive an
exponential solution, with the best example being a simple cosmological
constant.

Using the metric (\ref{FRW0metric0}) we examine whether the cosmological
equations (\ref{FR1})--(\ref{FR3}) admit solutions in
which the scale factor has an exponential dependence in time
of the form
\begin{eqnarray}
a(t)=e^{\frac{1}{2}H_{0}\, t}
,  \label{deSitteransantz}
\end{eqnarray}
with $H_{0}$ a constant. Inserting this in the first Friedmann equation (\ref%
{FR1}), and setting as usual $8\pi G=1$, we can easily see that the general
solution for the scalar field has a linear time dependence, thus we
depict our solution for the scalar field as
\begin{eqnarray}
\phi(t)=\phi_{1}\, t+\phi_{0},  \label{deSitterscalar}
\end{eqnarray}
where $\phi_{1},\phi_{0}$ are integration constants, which will be
determined in the following, in order for the full system of equations to be
satisfied. Inserting this expression for the scalar field in the
Klein-Gordon equation (\ref{FR3}) we deduce that it is satisfied if
\begin{eqnarray}
\beta=\frac{2\left(3H_{0}^{2}\zeta\phi_{1}^{2}-2 \right)}{3\,H_{0}^{2}}.
\end{eqnarray}
Substituting back to the Friedmann equations (\ref{FR1}) and (\ref{FR2}), we
obtain
\begin{eqnarray}
\zeta=\frac{2 \left(3H_{ 0}^{2}+4\phi_{1}^{2}
\right)}{3H_{0}^{2}\phi_{1}^{4}}.
\end{eqnarray}
Hence $\beta$ and $\zeta$ are both positive, namely
\begin{eqnarray}
\beta&=&\frac{4}{H_{0}^{2}}+\frac{1}{\phi_{1}^{2}}  \label{bsol} \\
\zeta&=&\frac{2}{\phi_{1}^{4}}+\frac{8}{3\,H_{0}^{2}\,\phi_{1}^{2}}.
\label{zsol}
\end{eqnarray}

In summary, we can see that the scenario at hand easily admits de-Sitter solutions, that
can
describe the inflationary epoch of the
Universe. In particular, for a given set of coupling parameters $\beta$ and $%
\zeta$, one obtains the de-Sitter solution (\ref{deSitteransantz}), with the
scalar field evolving as in Eq. (\ref{deSitterscalar}), where the solution
parameters $H_0$ and $\phi_1$ are determined by inverting Eqs. (\ref{bsol}) and (\ref%
{zsol}). The only requirement is the obtained $H_0$ and $\phi_1$ to be real
numbers, and this constrains the allowed $(\beta,z)$ parameter space. For
instance, note that if one of $\beta$ or $\zeta$ is zero, the system does
not admit an exponential solution unless there is a bare cosmological
constant. We stress that the above de-Sitter solution has been obtained
without considering a potential term, i.e. it is an effect of the
extended nonminimal derivative coupling terms considered in this work.

In general, at early times, where matter can be neglected, we can express the
deceleration parameter (\ref{deccparam})
using the Friedmann equations (\ref{FR1}) and (\ref{FR2}) as
\begin{equation}
q=-\frac{\dot{H}}{H^{2}}-1= \frac{1}{2}\left( 1+3w_{tot}\right) ,
\end{equation}%
where we have defined the ``total'' equation-of-state parameter of the Universe as
\begin{equation}
w_{tot}=\frac{\frac{\dot{\phi}^{2}}{2}-V(\phi )-\beta \frac{\dot{%
\phi}}{2}\Big[(3H^{2}+2\dot{H})\dot{\phi}+4H\ddot{\phi}\Big]+\zeta \frac{%
\dot{\phi}^{3}}{2}\Big[(2\dot{H}+3H^{2})\dot{\phi}+8H\,\ddot{\phi}\Big]}{\frac{\dot{\phi}^
{2}}{2}+V(\phi )+H^{2}\dot{%
\phi}^{2}\left( \frac{9}{2}\beta -\frac{15}{2}\zeta \,\dot{\phi}^{2}\right)}.
\end{equation}%
  Hence,
the condition for accelerated expansion can then be formulated as
$
1+3w_{tot}<0$,
or equivalently
\begin{equation}
\frac{\dot{\phi}\left\{ \dot{\phi}\left[ -3\beta \dot{H}+3\zeta \left( \dot{%
H}-H^{2}\right) \dot{\phi}^{2}+2\right] -6H\ddot{\phi}\left( \beta -2\zeta
\dot{\phi}^{2}\right) \right\} +4V(\phi )}{\dot{\phi}^{2}%
\left[ 3H^{2}\left( 3\beta -5\zeta \dot{\phi}^{2}\right) +1\right]  +2V(\phi )}<0,
\end{equation}
where for generality we have kept the potential term. Hence, one can use this requirement
in order
to find more general inflationary solutions, beyond the de-Sitter one.

\section{Late-time cosmology}\label{sec4}

In the present Section we investigate several cosmological models in the
framework of gravitational theories with an extended nonminimal derivative
coupling, focusing on the late-time evolution. In particular, we are interested in
studying the full Friedmann equations (\ref{FR1b}) and (\ref{FR2b}), i.e., considering
the
matter
sector as well, and we focus on important observables such as the dark-energy
equation-of-state parameter $w_{DE}$ defined in (\ref{wDE}), and the
dark-matter and dark-energy density parameters defined respectively as
\begin{eqnarray}
\Omega_m=\frac{8\pi G}{3 H^2}\rho_m,\qquad \Omega_{DE}=\frac{8\pi G}{3 H^2}%
\rho_{DE}.
\end{eqnarray}
Additionally, concerning the scalar potential we will consider three well-known
cases, namely the exponential potential
\cite{Copeland:1997et,Ferreira:1997au,Chen:2008ft,exsol1,exsol2}:
\begin{equation}
\label{exppot}
V(\phi )=\Lambda_0 e^{-\mu \phi},
\end{equation}
with $\Lambda _0$ and $\mu $  constants,
the power-law potential
\cite{Abramo:2003cp,Saridakis:2009pj}:
\begin{eqnarray}
V(\phi)=V_0 \phi^n,
\label{powerpot}
\end{eqnarray}
with $V_0$ and $n$ constants,
and the Higgs potential \cite{Lyth:1995hj}:
\begin{equation}
V(\phi )=V_{0}+\frac{1}{2}M^{2}\phi ^{2}+\frac{\lambda }{4}\phi ^{4},
\label{Higgspot}
\end{equation}%
where $V_{0}$ is a constant, while the constant $M^{2}<0$ may be related to
the mass of the Higgs boson by the relation $m_{H}=V^{\prime \prime }(v)$,
where $v^{2}=-M^{2}/\lambda $ gives the minimum of the potential.

In general, in the above cases, and in the presence of matter, analytical solutions are
impossible to be extracted, and thus we resort to numerical elaboration of the
cosmological equations. Thus, we evolve the equations using as independent
variable the redshift $z$ defined in (\ref{redshoftfedin}).

In order to perform a numerical elaboration of the above cosmological equations, it
proves convenient to re-write them in
dimensionless way. In particular, we introduce the dimensionless variables $\left(
\tau ,h,\Phi ,v\left( \Phi \right) ,\beta
_{0},\zeta _{0},r,P\right) $, defined as
\begin{eqnarray}  \label{var}
\tau =H_{0}t, \qquad  H=H_{0}h, \qquad  \phi =\sqrt{\frac{6}{8\pi G}}\Phi ,
\qquad  v\left( \Phi
\right) =\frac{8\pi G}{3H_{0}^{2}}V\left( \Phi \right) ,
     \\
\beta =\frac{\beta
_{0}}{9H_{0}^{2}}, \qquad \zeta =\frac{4\pi G}{45H_{0}^{4}}\zeta _{0},
\qquad  \rho _{m}=\frac{3H_{0}^{2}}{8\pi G}r, \qquad  p_{m}=\frac{3H_{0}^{2}}{8\pi G}P.
\end{eqnarray}
Using these new variables, the generalized Friedmann equations   (\ref{FR1}),
(\ref{FR2}), the scalar field equation  (\ref{FR3}), as well as the matter
conservation equation (\ref{rhoevol}), take the form
\begin{equation}
\frac{da}{d\tau }=\sqrt{\frac{v\left( \Phi \right) +r+\left( d\Phi /d\tau
\right) ^{2}}{1-\beta _{0}\left( d\Phi /d\tau \right) ^{2}+\zeta _{0}\left(
d\Phi /d\tau \right) ^{4}}}a\,,  \label{nd1}
\end{equation}
\begin{eqnarray}
2\frac{dh}{d\tau }+3h^{2}+3\left[ P-v\left( \Phi \right) +\left( \frac{d\Phi
}{d\tau }\right) ^{2}\right] -\frac{\beta _{0}}{3}\frac{d\Phi }{d\tau }\left[
\left( 2\frac{dh}{d\tau }+3h^{2}\right) \frac{d\Phi }{d\tau }+4h\frac{%
d^{2}\Phi }{d\tau ^{2}}\right]
    \nonumber  \\
+\frac{\zeta _{0}}{5}\left( \frac{d\Phi }{%
d\tau }\right) ^{3}\left[ \left( 2\frac{dh}{d\tau }+3h^{2}\right) \frac{%
d\Phi }{d\tau }+8h\frac{d^{2}\Phi }{d\tau ^{2}}\right] =0 \,,  \label{nd2}
\end{eqnarray}
\begin{equation}
\frac{d^{2}\Phi }{d\tau ^{2}}+3h\frac{d\Phi }{d\tau }+\frac{1}{2}\frac{%
dv\left( \Phi \right) }{d\Phi }+\frac{\beta _{0}}{3}h\left[ \left( 2\frac{dh%
}{d\tau }+3h^{2}\right) \frac{d\Phi }{d\tau }+h\frac{d^{2}\Phi }{d\tau ^{2}}%
\right] -\frac{2}{5}\zeta _{0}h\left( \frac{d\Phi }{d\tau }\right) ^{2}\left[
\left( 2\frac{dh}{d\tau }+3h^{2}\right) \frac{d\Phi }{d\tau }+3h\frac{%
d^{2}\Phi }{d\tau ^{2}}\right] =0 \,,  \label{nd3}
\end{equation}
\begin{equation}  \label{nd4}
\frac{dr}{d\tau}+3h\left(r+P\right)=0 \,,
\end{equation}
respectively.
 Therefore, after solving Eqs.~(\ref{nd2}) and (\ref{nd3}) for $dh/d\tau $ and $%
d^{2}\Phi /d\tau ^{2}$,   the cosmological field equations take the form
\begin{equation}
\frac{d\Phi }{d\tau }=\Pi ,
\label{d0}
\end{equation}
\begin{equation}
\frac{da}{d\tau }=\sqrt{\frac{v\left( \Phi \right) +r+\Pi ^{2}}{1-\beta
_{0}\Pi ^{2}+\zeta _{0}\Pi ^{4}}}a,  \label{d1}
\end{equation}%
\begin{eqnarray}
\frac{dh}{d\tau } &=&\frac{1}{2\left\{ h^{2}\left[ 5\beta _{0}+\left( 5\beta
_{0}^{2}-18\zeta _{0}\right) \Pi ^{2}-9\beta _{0}\zeta _{0}\Pi ^{4}+6\zeta
_{0}^{2}\Pi ^{6}\right] -5\beta _{0}\Pi ^{2}+3\zeta _{0}\Pi ^{4}+15\right\} }
     \nonumber  \\
&&\times
\Big\{-3\Pi ^{2}
\left\{2h^{2}[10\beta _{0}-9\zeta _{0}P+  9\zeta _{0}v(\Phi
)]+\left( 5\beta _{0}^{2}-18\zeta _{0}\right) h^{4}+15%
\right\}
-15\left( \beta _{0}h^{2}+3\right) \left[ h^{2}+P-v(\Phi )\right]
\nonumber
\\
&&
+9\zeta _{0}h^{2}\Pi ^{4}\left( 3\beta _{0}h^{2}+13\right) -18\zeta
_{0}^{2}h^{4}\Pi ^{6}+  2h\Pi \left( 6\zeta _{0}\Pi ^{2}-5\beta _{0}\right) v^{\prime
}(\Phi )%
\Big\}, \label{d2}
\end{eqnarray}%
\begin{equation}
\frac{d\Pi }{d\tau }=\frac{\left( 5\beta _{0}\Pi ^{2}-3\zeta _{0}\Pi
^{4}-15\right) v^{\prime }(\Phi )-6h\Pi \left[ -5\beta _{0}P-2\Pi
^{2}(5\beta _{0}-3\zeta _{0}P+3\zeta _{0}v(\Phi ))+9\zeta _{0}\Pi
^{4}+5\beta _{0}v(\Phi )+15\right] }{2\left\{ h^{2}\left[ 5\beta _{0}+\left(
5\beta _{0}^{2}-18\zeta _{0}\right) \Pi ^{2}-9\beta _{0}\zeta _{0}\Pi
^{4}+6\zeta _{0}^{2}\Pi ^{6}\right] -5\beta _{0}\Pi ^{2}+3\zeta _{0}\Pi
^{4}+15\right\} },  \label{d3}
\end{equation}%
which must be solved together with Eq.~(\ref{nd4}) after the equation of
state $P=P(r)$ of matter has been imposed. The initial conditions for
the system (\ref{d0})-(\ref{d3}) are $a\left( \tau _{0}\right) =a_{0}$, $%
h\left( \tau _{0}\right) =h_{0}$, $\Phi \left( \tau _{0}\right) =\Phi _{0}$,
and $\Pi \left( \tau _{0}\right) =\Pi _{0}$, respectively. Furthermore, in terms of the
dimensionless variables the deceleration
parameter  (\ref{deccparam})
becomes
\begin{equation}
q=\frac{d}{d\tau }\left( \frac{1}{h}\right) -1,
\end{equation}
while the dark-energy equation-of-state parameter reads as
\begin{equation}
w_{DE}=\frac{\Pi ^{2}-v\left( \Phi \right) -\frac{\beta _{0}}{9}\Pi \left[
\left( 2\frac{dh}{d\tau }+3h^{2}\right) \Pi +4h\frac{d\Pi }{d\tau }\right] +%
\frac{\zeta _{0}}{15}\Pi ^{3}\left[ \left( 2\frac{dh}{d\tau }+3h^{2}\right)
\Pi +8h\frac{d\Pi }{d\tau }\right] }{\Pi ^{2}\left\{ 1+h^{2}\left[ \beta
_{0}-\zeta _{0}\Pi ^{2}\right] \right\} +v\left( \Phi \right) }.
\end{equation}
Finally,   the dimensionless time-redshift relation (\ref{timeredshiftrel}) becomes
\begin{equation}
\frac{d}{d\tau}=-(1+z)h(z)\frac{d}{dz}.
\end{equation}

\subsection{Exponential potential}

Let us start the analysis by considering the exponential potential (\ref{exppot}), namely
\begin{equation}
\label{exppot2}
V(\phi )=\Lambda_0\exp \left(-\mu \phi\right),
\end{equation}
with $\Lambda _0$ and $\mu $ the potential parameters.   In terms of the dimensionless
variables introduced in (\ref{var}), the above exponential
potential takes the form
\begin{equation}  \label{ep}
v\left(\Phi \right)=\lambda _0\exp \left(-\mu _0\Phi \right),
\end{equation}
where
\begin{equation}
\lambda _0=\frac{8\pi G}{3H_0^2}\Lambda _0, \ \ \ \mu _0=\sqrt{\frac{6}{8\pi G}}%
\mu.
\end{equation}


In the following we consider the time evolution of a dust matter fluid, namely we assume
that  $P=0$, and hence the redshift dependence of
the auxiliary matter energy density is given by $r=(1+z)^3$. In this case, the
dimensionless cosmological equations
(\ref{d0}), (\ref{d2}) and (\ref{d3}) become
\begin{equation}  \label{e1}
\frac{d\Phi }{dz}=-\frac{\Pi }{(1+z)h},
\end{equation}%
\begin{eqnarray}  \label{e2}
\frac{dh}{dz} &=&
-
\left\{
\frac{(1+z)^{-1}h^{-1}}{2\left\{ h^{2}\left[ 5\beta
_{0}+\left( 5\beta _{0}^{2}-18\zeta _{0}\right) \Pi ^{2}-9\beta _{0}\zeta
_{0}\Pi ^{4}+6\zeta _{0}^{2}\Pi ^{6}\right] -5\beta _{0}\Pi ^{2}+3\zeta
_{0}\Pi ^{4}+15\right\} }   \right\}
   \nonumber   \\
&&\times \Bigg\{-3\Pi ^{2}\Bigg[2h^{2}(10\beta _{0}+
9\zeta _{0}\lambda _{0}e^{-\mu _{0}\Phi }+\left( 5\beta _{0}^{2}-18\zeta
_{0}\right) h^{4}+15\Bigg] -15\left( \beta _{0}h^{2}+3\right) \left(
h^{2}-\lambda _{0}e^{-\mu _{0}\Phi }\right)
    \nonumber \\
&& \ \ \ \ \
+9\zeta _{0}h^{2}\Pi ^{4}\left(
3\beta _{0}h^{2}+13\right) -18\zeta _{0}^{2}h^{4}\Pi ^{6}-  2h\Pi \left( 6\zeta _{0}\Pi
^{2}-5\beta
_{0}\right) \mu _{0}\lambda
_{0}e^{-\mu _{0}\Phi }\Bigg\},
\end{eqnarray}%
\begin{equation}  \label{e3}
\frac{d\Pi }{dz}=\frac{\left( 5\beta _{0}\Pi ^{2}-3\zeta _{0}\Pi
^{4}-15\right) \mu _{0}\lambda _{0}e^{-\mu _{0}\Phi }+6h\Pi \left[ -2\Pi
^{2}(5\beta _{0}+3\zeta _{0}\lambda _{0}e^{-\mu _{0}\Phi })+9\zeta _{0}\Pi
^{4}+5\beta _{0}\mu _{0}\lambda _{0}e^{-\mu _{0}\Phi }+15\right] }{%
2(1+z)h\left\{ h^{2}\left[ 5\beta _{0}+\left( 5\beta _{0}^{2}-18\zeta
_{0}\right) \Pi ^{2}-9\beta _{0}\zeta _{0}\Pi ^{4}+6\zeta _{0}^{2}\Pi ^{6}%
\right] -5\beta _{0}\Pi ^{2}+3\zeta _{0}\Pi ^{4}+15\right\} },
\end{equation}
respectively. Additionally, the parameter of the dark energy equation of state
(\ref{wDE}) reads as
\begin{eqnarray}
&&\!\!\!\!\!\!\!\!\!\!
w_{DE}=
\left\{\Pi ^{2}\left[ 1+h^{2}\left( \beta _{0}-\zeta _{0}\Pi ^{2}\right)
\right] +\lambda _{0}e^{-\mu _{0}\Phi }\right\}^{-1}
\Bigg\{\Pi ^{2}-\lambda _{0}e^{-\mu _{0}\Phi } -\frac{\beta _{0}}{9}\Pi %
\left\{ \left[ -2\left( 1+z\right) h\frac{dh}{dz}+3h^{2}\right] \Pi
-4(1+z)h^{2}\frac{d\Pi }{dz}\right\}
    \nonumber  \\
&&
\ \ \ \ \ \ \ \ \ \ \ \ \ \ \ \ \ \ \ \ \ \ \ \ \ \ \ \ \ \ \ \ \ \ \ \ \ \ \ \ \ \ \ \ \
\ \ \ \ \ \ \ \ \ \ \ \ \ \ \
+\frac{\zeta _{0}}{15}\Pi ^{3}
\left\{
\left[ -2(1+z)h\frac{dh}{dz}+3h^{2}\right] \Pi -8(1+z)h^{2}\frac{d\Pi }{dz}%
\right\}\Bigg\} .
\end{eqnarray}
A crucial observation is that according to the above expression, $w_{DE}$ could  acquire
values
below $-1$ too, and thus the phantom regime can be exhibited. This is an advantage of the
scenario
at hand, since such a behavior is obtained although the scalar field is canonical, that
it
is a
pure result of the extended, gravitational couplings.

In order to study the cosmological evolution of the dust Universe in the presence of the
exponential
 potential we integrate the system of Eqs.~(\ref{e1})-(\ref{e3}) numerically. We choose
the
potential parameters as  $\lambda_0=0.36$, and $\mu_0=-1.05$, while for the initial
conditions we
set  $h(0)=1$, $\Phi(0)=1$, and $\Pi (0)=0.1$. We are interested in studying the effect
of
the
parameters $\beta _0$ and $\zeta _0$, that determine the novel, extended nonminimal
derivative
coupling, on the cosmological evolution, restricting the analysis at late times, i.e., at
the
redshift range $0\leq z \leq 1$.

\begin{figure*}[ht]
\centering
\includegraphics[width=8.15cm]{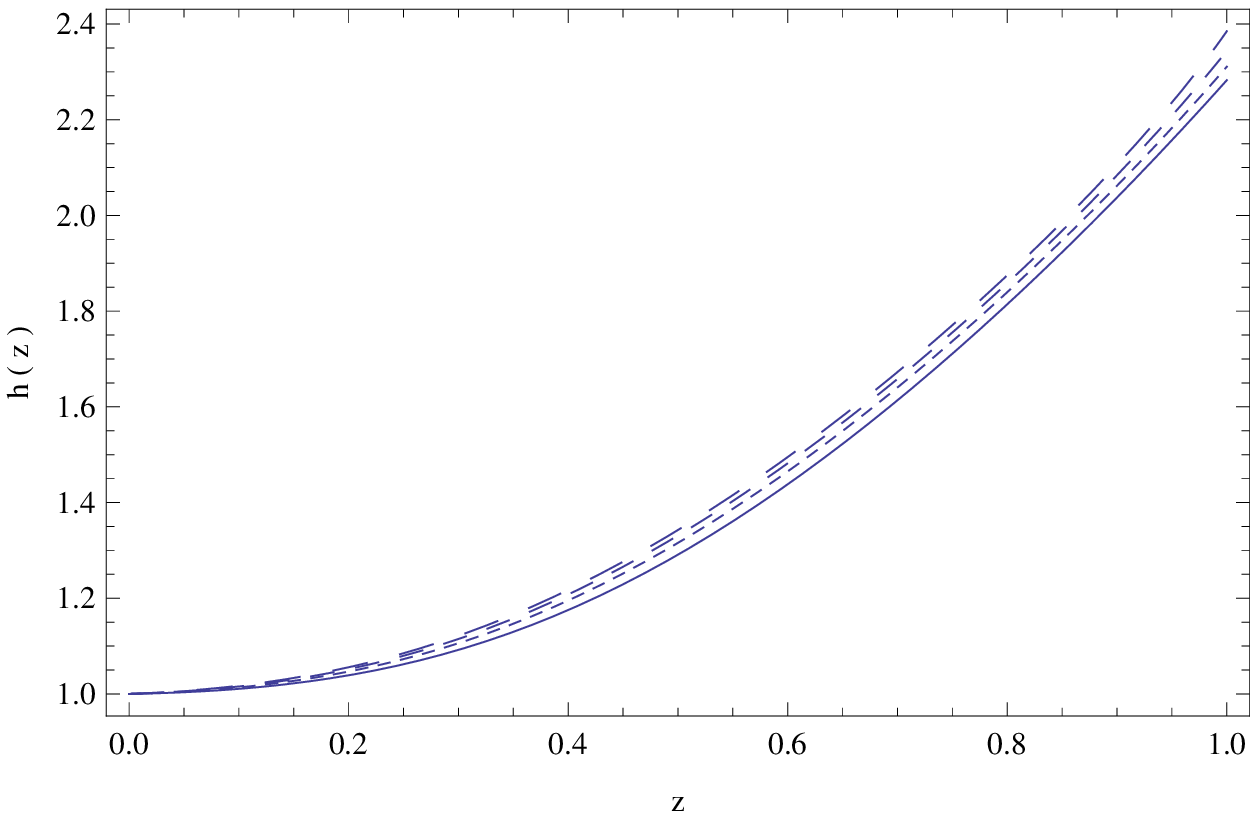} %
   \hspace{0.5cm}
\includegraphics[width=8.15cm]{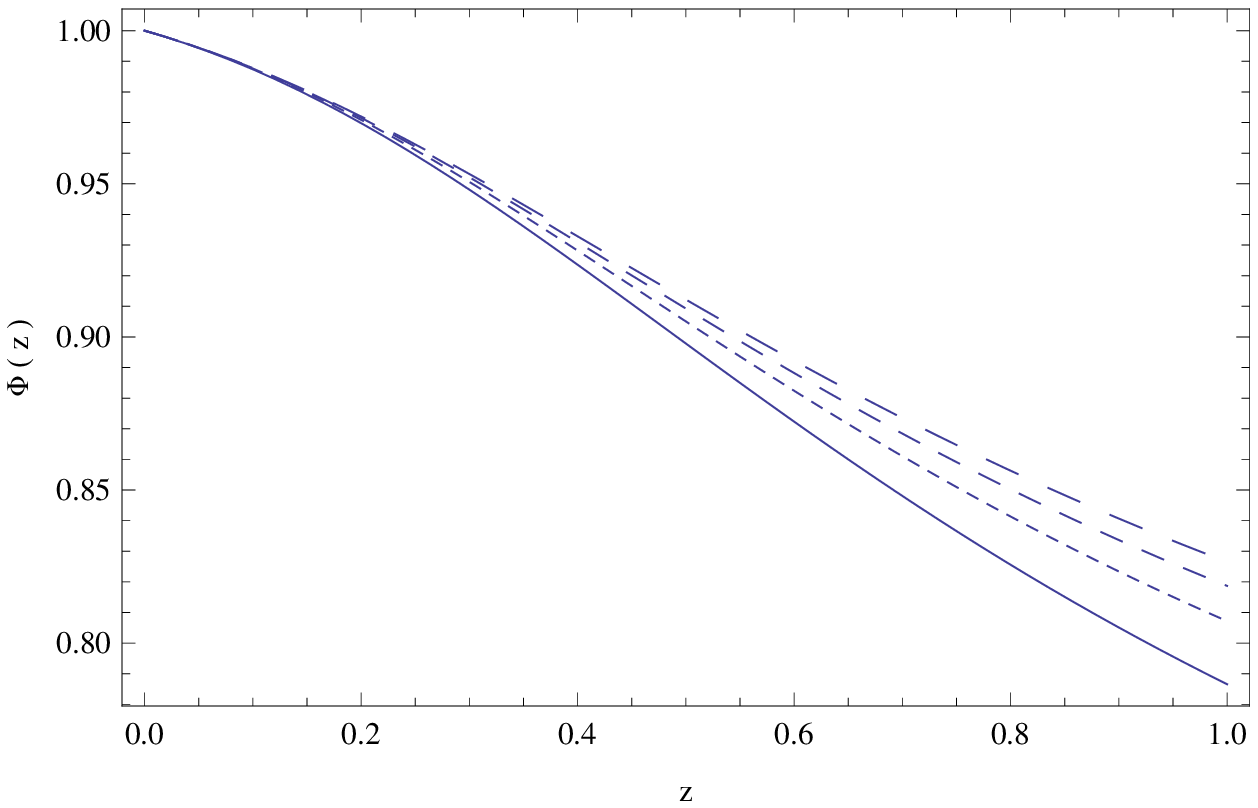} %
\includegraphics[width=8.15cm]{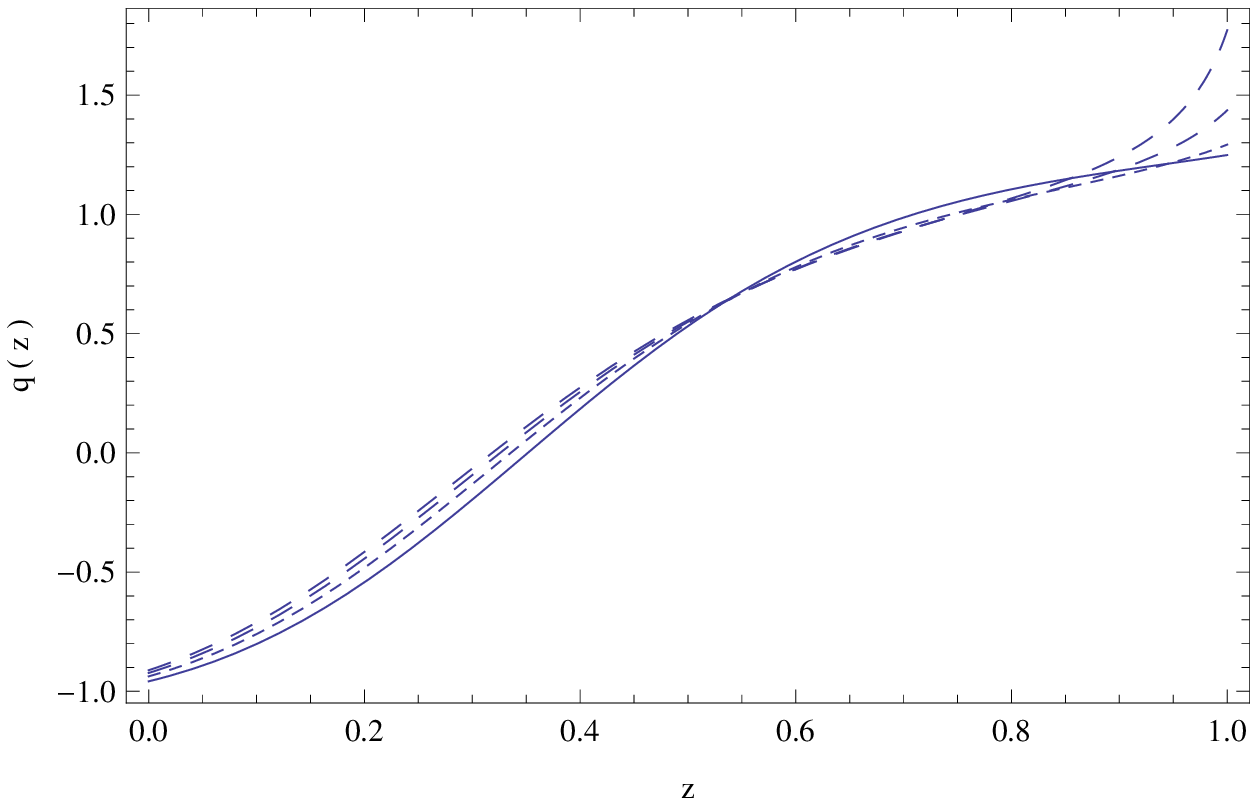}
   \hspace{0.5cm}
\includegraphics[width=8.15cm]{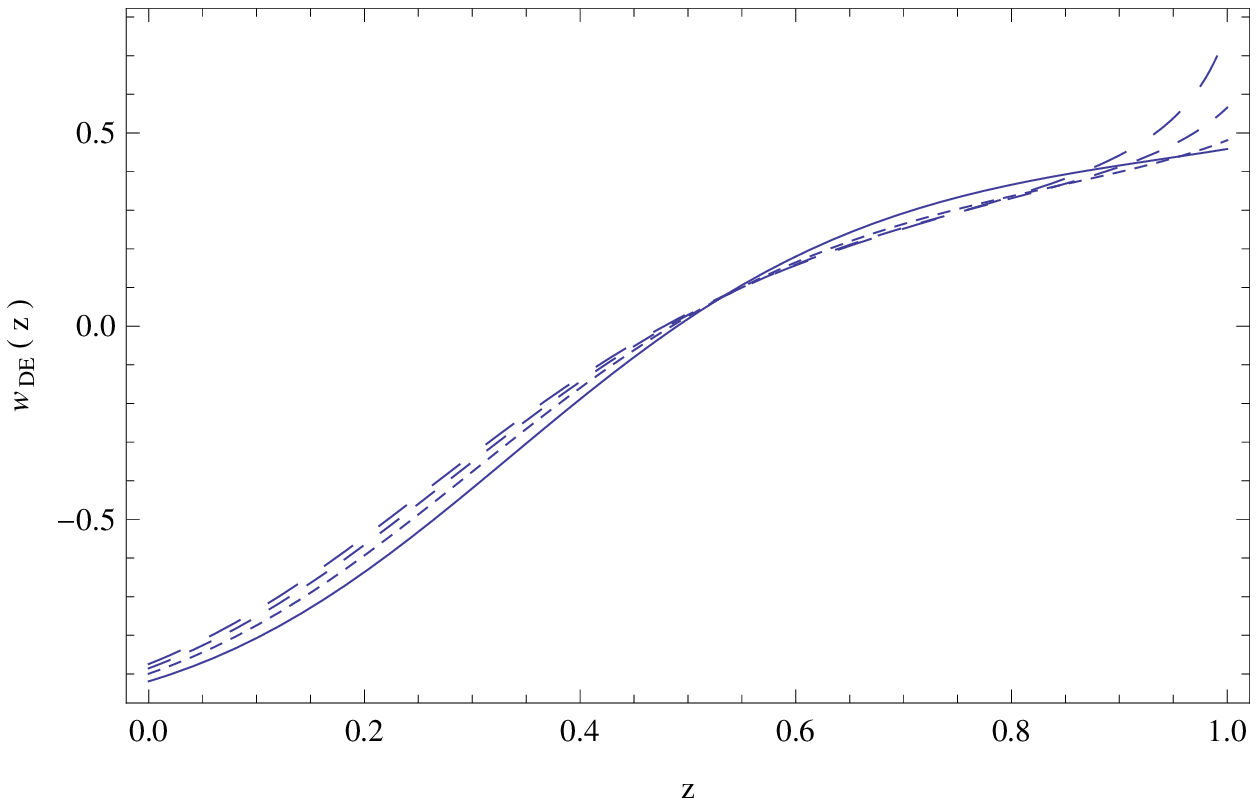}
\caption{ {\it{Evolution of the dimensionless Hubble function (top left), of the
dimensionless field (top right), of the deceleration parameter (bottom left), and of the
dark-energy equation-of-state parameter (bottom right), as a function of the redshift,
for
cosmology with extended nonminimal derivative coupling in the case of the exponential
potential (\ref{exppot2}) and for dust matter. The initial conditions have been chosen as
$h(0)=1$, $\Phi(0)=1$, and $\Pi (0)=0.1$, while the parameters of the exponential
potential have been fixed as $\mu_0=-1.05$ and $\lambda_0=0.36$. Concerning
the coupling parameters $\beta _0$ and $\zeta _0$, we choose: $\beta _0=1.90$ and $\zeta
_0=1$ (solid curve), $\beta _0=2.68$ and $\zeta _0=2$ (dotted curve), $\beta _0=3.29$ and
$\zeta _0=3$ (short dashed curve) and $\beta _0=3.79$ and $\zeta _0=4$ (dashed curve),
respectively.  }}}
\label{fig3}
\end{figure*}

In Fig.~\ref{fig3}, we depict the evolution of the Hubble function, of the scalar field,
of the
deceleration parameter, and of the dark-energy equation-of-state parameter, in terms of
the
redshift, for various values of $\beta _0$ and $\zeta _0$.
As we can see, the Hubble function, represented in the top left figure, is a
monotonically increasing function of the redshift, indicating an expansionary
cosmological
evolution, while in the redshift range $0\leq z\leq 0.1$ it remains almost constant. Its
variation is not affected significantly by the change of the numerical values of the
coupling parameters. Moreover, the dimensionless scalar field, depicted in the top right
figure, is a monotonically decreasing function of the redshift, and therefore an
increasing function of the cosmological time. Note that the scalar field behavior is
strongly affected by variations of the numerical values of $\beta_0$ and $\zeta _0$. The
deceleration parameter, shown in the bottom left figure, starts with high positive
values,
however acceleration arises quickly, with $q$ crossing zero at a redshift $z\approx
0.35$.
The variations of the coupling parameters have a small influence on the deceleration
parameter behavior. Finally, the parameter $w_{DE}$ of the dark energy equation of state,
presented in the bottom right figure, starts with values of the order of $w_{DE}\approx
0.4-0.5$ at $z=1$, it reaches zero at $z\approx 0.5$, and tends to $-1$ at $z=0$,
indicating that the dark energy sector behaves like a cosmological constant at present.
Similarly to the case of the dimensionless Hubble parameter, the changes in the numerical
values of the parameters $\beta _0$ and $\zeta _0$ have small influence on $w_{DE}$
evolution.

\subsection{Power-law potential}

Let us now investigate the cosmological evolution in the extended
nonminimal derivative coupling gravitational theory in the presence of a
simple power-law potential of the form (\ref{powerpot}), namely
\begin{equation}
\label{powerpot2}
V(\phi)=V_0\phi ^n,
\end{equation}
where $V_0$ and $n$ are constants. Hence, using the dimensionless
variables introduced in (\ref{var}), the dimensionless form of the power-law
potential writes as
\begin{equation}
v\left(\Phi \right)=u_0\Phi ^n,
\end{equation}
where
\begin{equation}
u_0=\frac{6^{n/2}}{3H_0^2}\left(8\pi G\right)^{1-n/2}V_0.
\end{equation}


In the case of the dust Universe, namely imposing that $P=0$,      the
dimensionless cosmological equations
(\ref{d0}), (\ref{d2}) and (\ref{d3}) become
\begin{equation}\label{pp1}
\frac{d\Phi }{dz}=-\frac{\Pi }{(1+z)h},
\end{equation}
\begin{eqnarray}\label{pp2}
\frac{dh}{dz} &=&
-
\left\{\frac{(1+z)^{-1}h^{-1}}{2\left\{ h^{2}\left[ 5\beta
_{0}+\left( 5\beta _{0}^{2}-18\zeta _{0}\right) \Pi ^{2}-9\beta _{0}\zeta
_{0}\Pi ^{4}+6\zeta _{0}^{2}\Pi ^{6}\right] -5\beta _{0}\Pi ^{2}+3\zeta
_{0}\Pi ^{4}+15\right\} }\right\}
   \nonumber  \\
&& \times \Bigg\{-3\Pi ^{2}\Big[2h^{2}(10\beta _{0}+  9\zeta _{0}u_{0}\Phi ^{n})+\left(
5\beta _{0}
^{2}-18\zeta _{0}\right)
h^{4}+15\Big] -15\left( \beta _{0}h^{2}+3\right) \left( h^{2}-u_{0}\Phi
^{n}\right)
       \nonumber \\
&& \ \ \ \ \
+9\zeta _{0}h^{2}\Pi ^{4}\left( 3\beta _{0}h^{2}+13\right)
-18\zeta _{0}^{2}h^{4}\Pi ^{6}+  2h\Pi \left( 6\zeta _{0}\Pi ^{2}-5\beta _{0}\right)
nu_{0}\Phi ^{n-
1}%
\Bigg\},
\end{eqnarray}%
\begin{equation}\label{pp3}
\frac{d\Pi }{dz}=-\frac{\left( 5\beta _{0}\Pi ^{2}-3\zeta _{0}\Pi
^{4}-15\right) v^{\prime }(\Phi )
-6h\Pi \left\{ -2\Pi ^{2}[5\beta _{0}+3\zeta
_{0}v(\Phi )]+9\zeta _{0}\Pi ^{4}+5\beta _{0}u_{0}\Phi ^{n}+15\right\} }{%
2(1+z)h\left\{ h^{2}\left[ 5\beta _{0}+\left( 5\beta _{0}^{2}-18\zeta
_{0}\right) \Pi ^{2}-9\beta _{0}\zeta _{0}\Pi ^{4}+6\zeta _{0}^{2}\Pi ^{6}%
\right] -5\beta _{0}\Pi ^{2}+3\zeta _{0}\Pi ^{4}+15\right\} },
\end{equation}
respectively. Additionally, the dark-energy equation-of-state parameter (\ref{wDE})
writes
as
\begin{eqnarray}
&&\!\!\!\!\!\!\!\!\!\!
w_{DE}=
\left\{\Pi ^{2}\left[1+h^{2}\left( \beta _{0}-\zeta _{0}\Pi ^{2}\right) \right]
+u_{0}\Phi
^{n}\right\}^{-1}
\Bigg\{\Pi ^{2}-u_{0}\Phi ^{n}-\frac{\beta _{0}}{9}\Pi
\left\{ \left[
-2\left( 1+z\right) h\frac{dh}{dz}+3h^{2}\right] \Pi -4(1+z)h^{2}\frac{d\Pi
}{dz}\right\}
   \nonumber  \\
   &&
\ \ \ \ \ \ \ \ \ \ \ \ \ \ \ \ \ \ \ \ \ \ \ \ \ \ \ \ \ \ \ \ \ \ \ \ \ \ \ \ \ \ \ \ \
\ \ \ \ \ \ \ \ \ \ \ \,
+\frac{\zeta _{0}}{15}\Pi ^{3}\left\{ \left[
-2(1+z)h\frac{dh}{dz%
}+3h^{2}\right] \Pi -8(1+z)h^{2}\frac{d\Pi }{dz}\right\} \Bigg\}
   .\label{pp4}
\end{eqnarray}
Similarly to the exponential potential case, we can see that $w_{DE}$ can acquire values
of the phantom regime, which is an advantage of the scenario at hand.

In order to study the cosmological evolution of the dust Universe in the presence of the
power-law potential we integrate the system of Eqs.~(\ref{pp1})-(\ref{pp3}) numerically.
We choose the potential parameters as  $n=1/4$ and $u_0=1.05$, and for the initial
conditions we set $h(0)=1$, $\Phi(0)=1$, and $\Pi (0)=0.1$, and similarly to the
exponential potential of the previous subsection we restrict our analysis at late times,
namely, at the redshift range $0\leq z \leq 1$.

\begin{figure*}[ht]
\centering
\includegraphics[width=8.15cm]{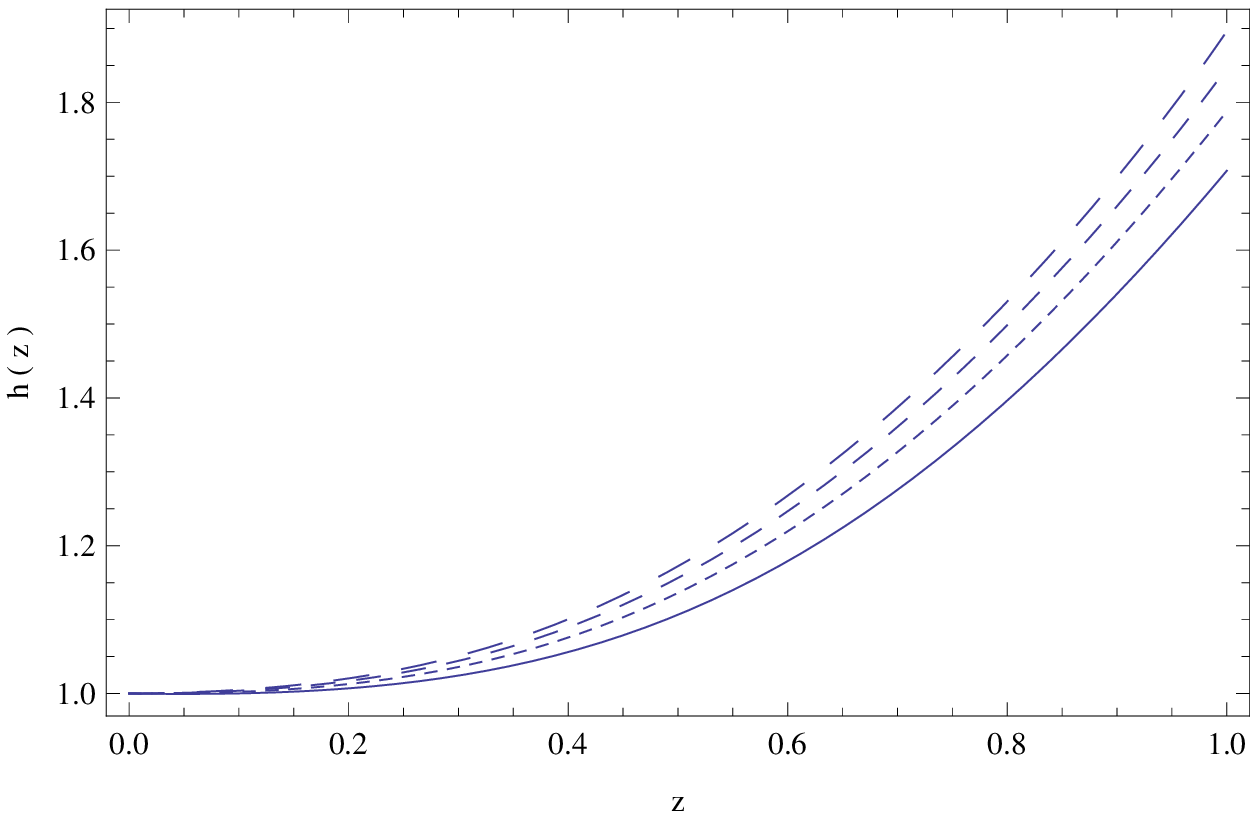} %
  \hspace{0.5cm}
\includegraphics[width=8.15cm]{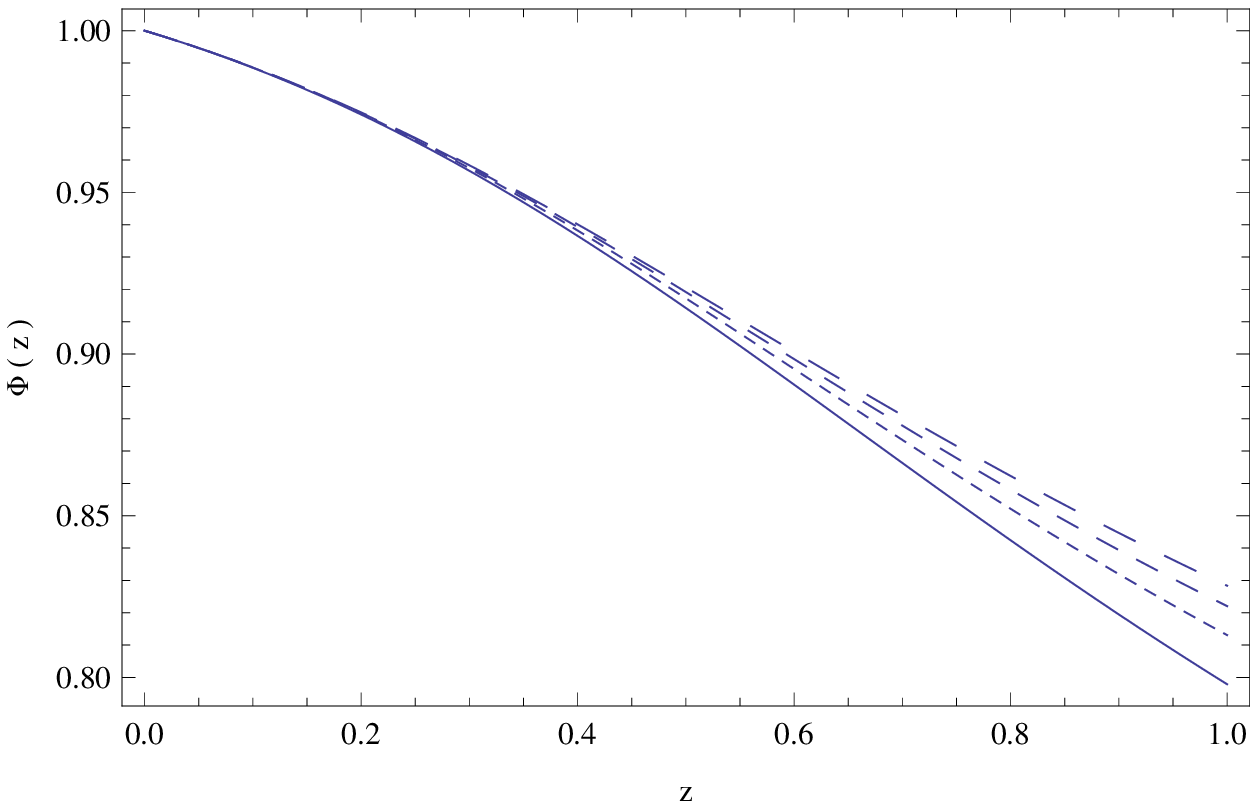} %
\includegraphics[width=8.15cm]{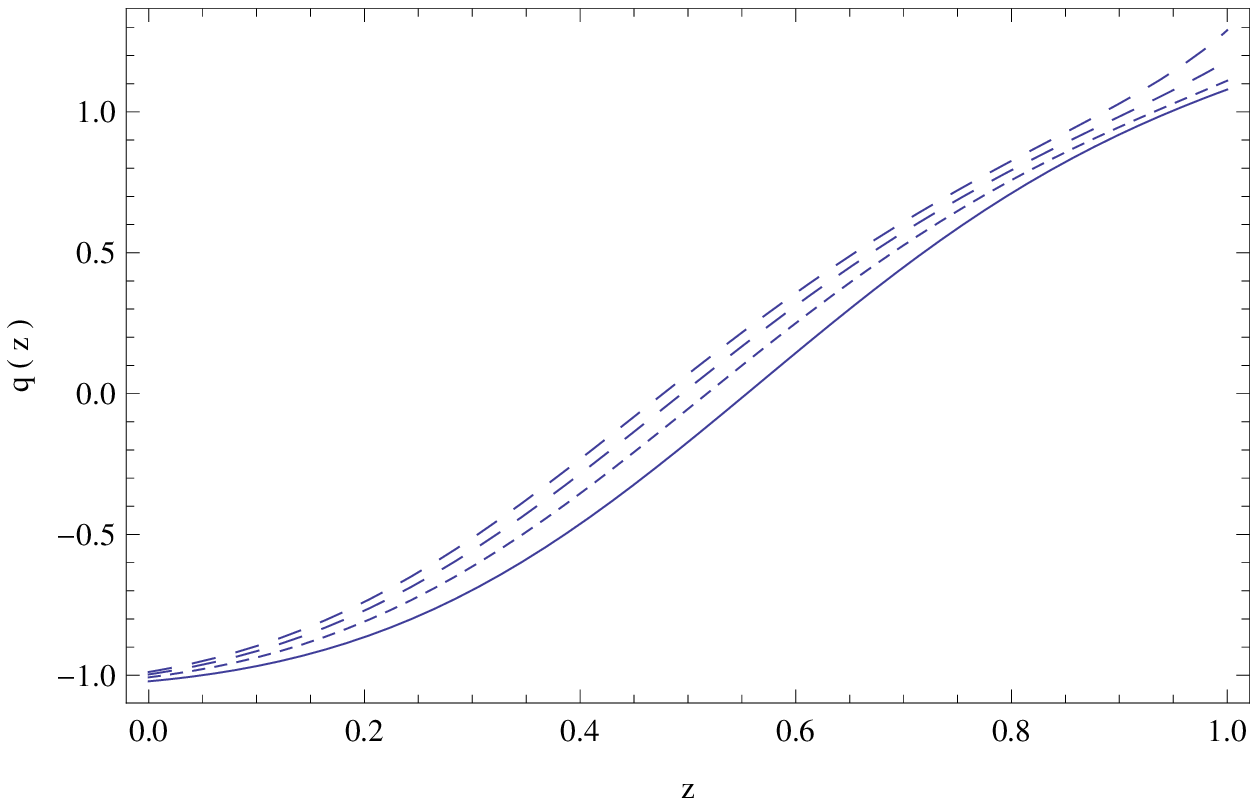}
   \hspace{0.5cm}
\includegraphics[width=8.15cm]{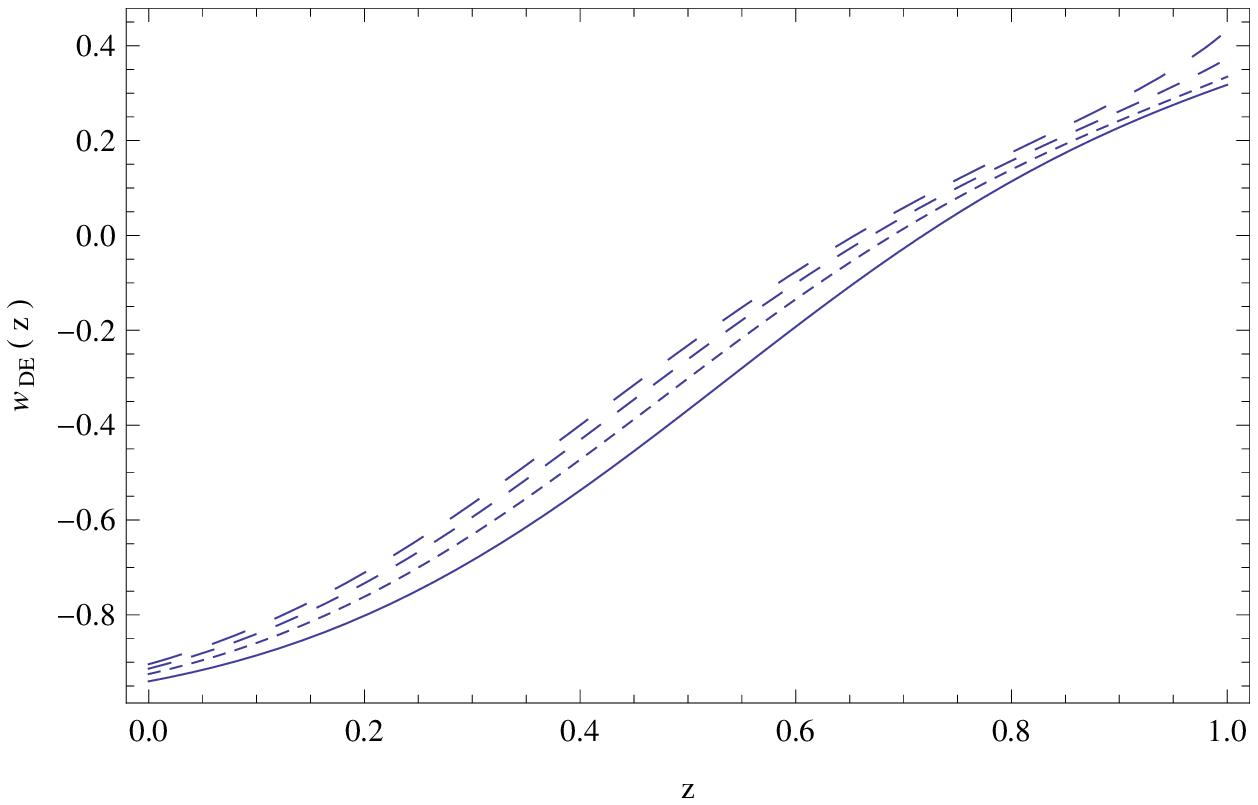}
\caption{
{\it{Evolution of the dimensionless Hubble function (top left), of the dimensionless field
(top right), of the deceleration parameter (bottom left), and of the dark-energy
equation-of-state parameter (bottom right), as a function of the redshift, for cosmology
with extended nonminimal derivative coupling in the case of the power-law potential
(\ref{powerpot2}) and for dust matter. The initial conditions have been chosen as
$h(0)=1$, $\Phi(0)=1$, and $\Pi (0)=0.1$, while the parameters of the
potential have been fixed as  $n=1/4$ and $u_0=1.05$. Concerning the coupling parameters
$\beta _0$ and $\zeta _0$, we choose: $\beta _0=1.90$ and $\zeta _0=1$ (solid curve),
$\beta _0=2.68$ and $\zeta _0=2$ (dotted curve), $\beta _0=3.29$ and $\zeta _0=3$ (short
dashed curve) and $\beta _0=3.79$ and $\zeta _0=4$ (dashed curve) respectively.  }} }
\label{fig4}
\end{figure*}

As in the previous case, in Fig.~\ref{fig4} we depict the evolution of the Hubble
function, of the
scalar field,
of the deceleration parameter, and of the dark-energy equation-of-state parameter, in
terms of the redshift, for various values of the coupling parameters $\beta _0$ and $\zeta
_0$.

The evolution of the Hubble function, presented in the top left graph, shows that the
Universe is expanding, with the Hubble function monotonically increasing with the
redshift, while at $z\approx 0.2$ and below the Hubble function becomes almost a
constant. The behavior of the Hubble function is relatively strongly affected by
changes in the values of the coupling parameters $\beta _0$ and $\zeta _0$, with the
effect being stronger at higher redshifts. The dimensionless scalar field $\Phi $,
depicted in the top right graph, is a monotonically decreasing function of the
redshift, and at high redshifts it also presents a strong dependence on the numerical
values of $\beta _0$ and $\zeta_0$. The deceleration parameter, presented in the bottom
left graph, has positive values of the order of $q=1$ at $z=1$, however acceleration is
obtained  at a redhift $z\approx 0.5$. Finally, the dark-energy equation-of-state
parameter, presented in the bottom right graph, has values of the order of $w_{DE}\approx
0.3-0.4$ at $z=1$, it reaches zero at redshifts $z\approx 0.6-0.7$, and it  tends to $-1$
at $z=0$, implying that the dark energy sector behaves like a cosmological constant at
present. We mention that both $q$ and $w_{DE}$ exhibit a relatively strong dependence
on the numerical values of $\beta _0$ and $\zeta_0$.

\subsection{Higgs potential}

As a final case we investigate the cosmological implications of theories with extended
nonminimal derivative coupling, in the presence of the Higgs potential
(\ref{Higgspot}), namely
\begin{equation}
V(\phi )=V_{0}+\frac{1}{2}M^{2}\phi ^{2}+\frac{\lambda }{4}\phi ^{4},
\label{Higgspot2}
\end{equation}%
where $V_{0}$ is a constant, and where the constant $M^{2}<0$ can be related to
the Higgs mass by the relation $m_{H}=V^{\prime \prime }(v)$,
with $v^{2}=-M^{2}/\lambda $   the minimum of the potential. Moreover, based on
the determination of $m_{H}$ from accelerator experiments one can infer for
the Higgs self-coupling constant a value of the order of $\lambda \approx
1/8 $ \cite{Higgs}. In terms of the dimensionless variables (\ref{var}) the above
Higgs-like potential becomes
\begin{equation}
v(\Phi)=v_0-\frac{1}{2}m^2\Phi ^2+\frac{1}{4}\Lambda \Phi ^4,
\end{equation}
where
\begin{equation}
v_0=\frac{8\pi G}{3H_0^2}V_0, \qquad m^2=2\left(\frac{M}{H_0}\right)^2, \qquad \Lambda =
\frac{12\lambda }{8\pi GH_0^2}.
\end{equation}
In the following paragraphs we study separately the cases of dust and radiation,
respectively.

\subsubsection{Cosmological evolution of a dust fluid}

In the case of the dust matter sector, namely for $P=0$, the dimensionless cosmological
equations (\ref{d0}), (\ref{d2}) and (\ref{d3}) become
\begin{equation}
\frac{d\Phi }{dz}=-\frac{\Pi }{(1+z)h},  \label{h0}
\end{equation}%
\begin{eqnarray}
\frac{dh}{dz} &=&-\left\{\frac{(1+z)^{-1}h^{-1}}{2\left\{ h^{2}\left[ 5\beta
_{0}+\left( 5\beta _{0}^{2}-18\zeta _{0}\right) \Pi ^{2}-9\beta _{0}\zeta
_{0}\Pi ^{4}+6\zeta _{0}^{2}\Pi ^{6}\right] -5\beta _{0}\Pi ^{2}+3\zeta
_{0}\Pi ^{4}+15\right\} }  \right\}
    \nonumber \\
&&\times\Bigg\{-3\Pi ^{2}
\Bigg\{2h^{2}
\left[10\beta _{0}+
9\zeta _{0}\left( v_{0}-\frac{1}{2}m^{2}\Phi ^{2}+\frac{1}{4}\Lambda \Phi
^{4}\right) \right]+\left( 5\beta _{0}^{2}-18\zeta _{0}\right) h^{4}+15\Bigg\}
  \nonumber  \label{h1} \\
&&\ \ \ \ \ \,
-15\left( \beta _{0}h^{2}+3\right) \left[ h^{2}-\left( v_{0}-\frac{1}{2}%
m^{2}\Phi ^{2}+\frac{1}{4}\Lambda \Phi ^{4}\right) \right] +
9\zeta _{0}h^{2}\Pi ^{4}\left( 3\beta _{0}h^{2}+13\right) -18\zeta_{0}^{2}h^{4}\Pi ^{6}
      \nonumber \\
&&\ \ \ \ \ \,
+2h\Pi \left( 6\zeta _{0}\Pi ^{2}-5\beta _{0}\right)
\left( -m^{2}\Phi +\Lambda \Phi ^{3}\right) \Bigg\}\,,
\end{eqnarray}%
\begin{figure*}[ht]
\centering
\includegraphics[width=8.15cm]{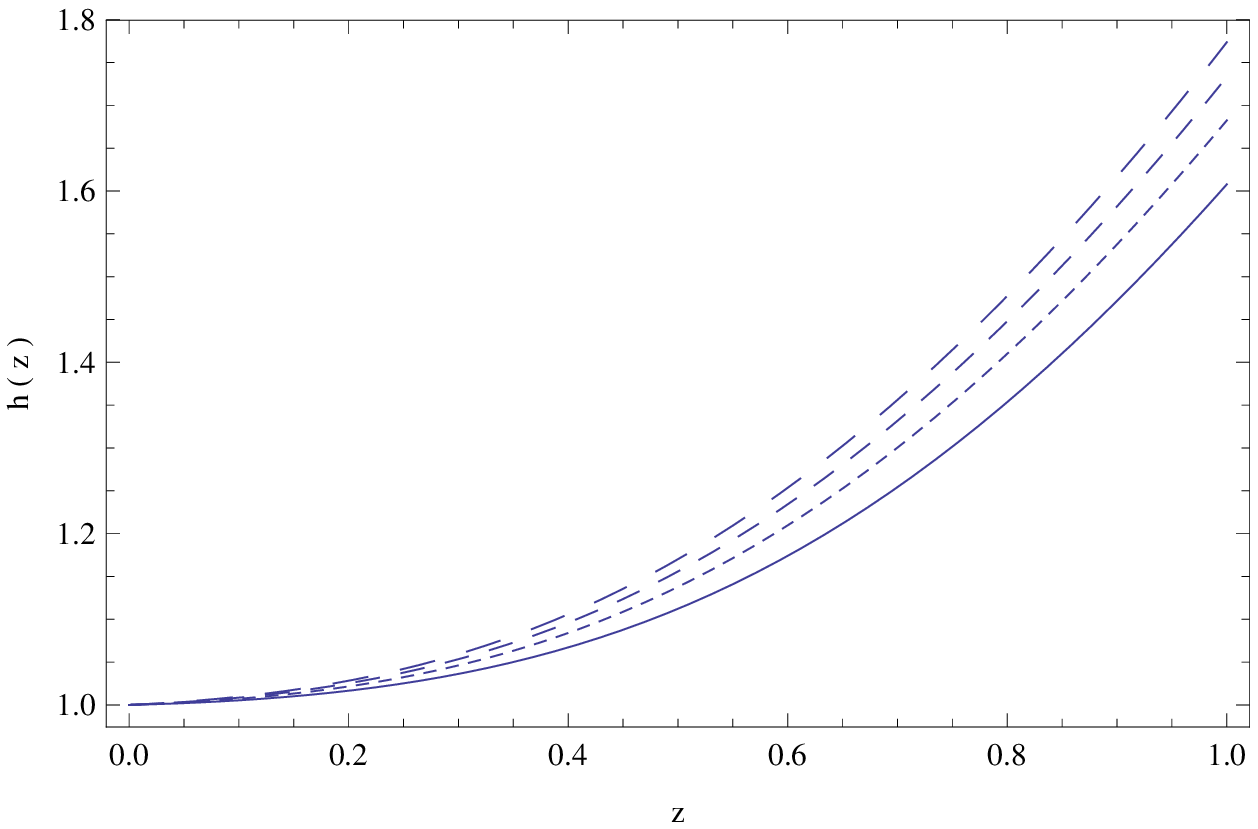} %
   \hspace{0.5cm}
\includegraphics[width=8.15cm]{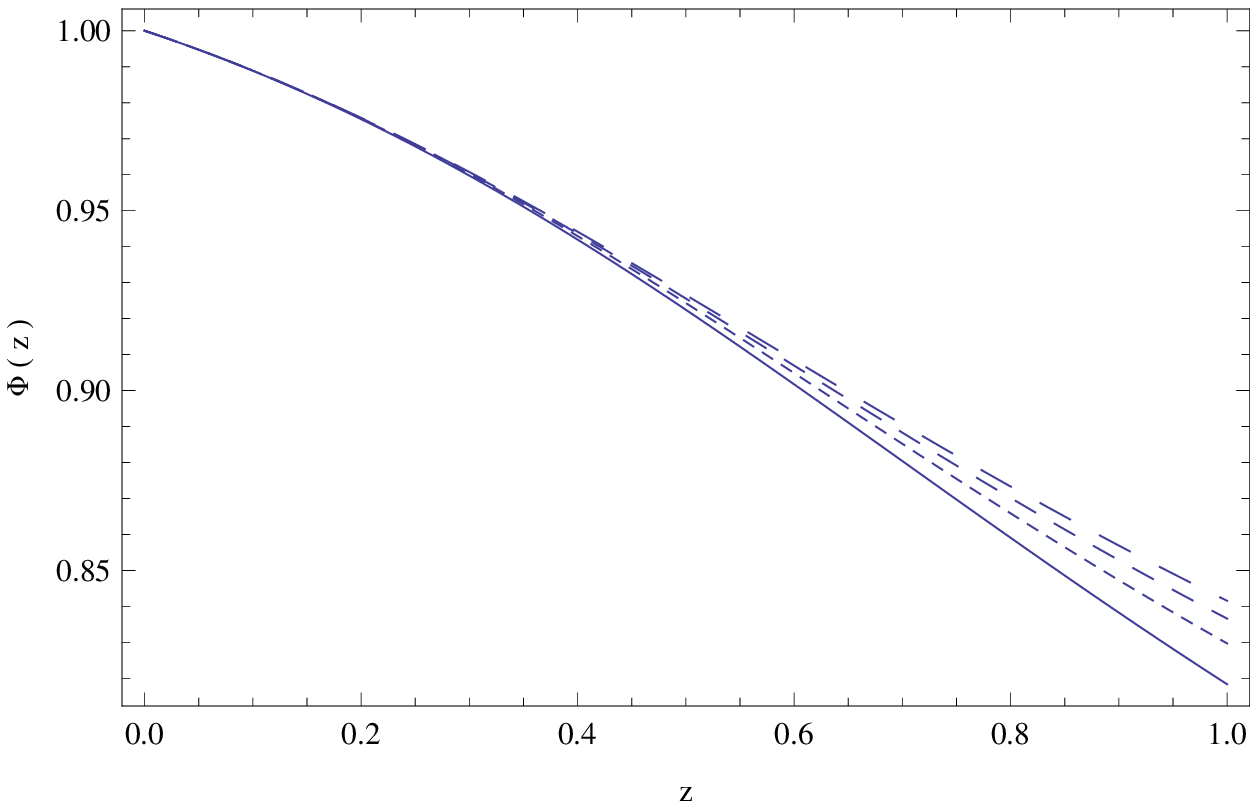} %
\includegraphics[width=8.15cm]{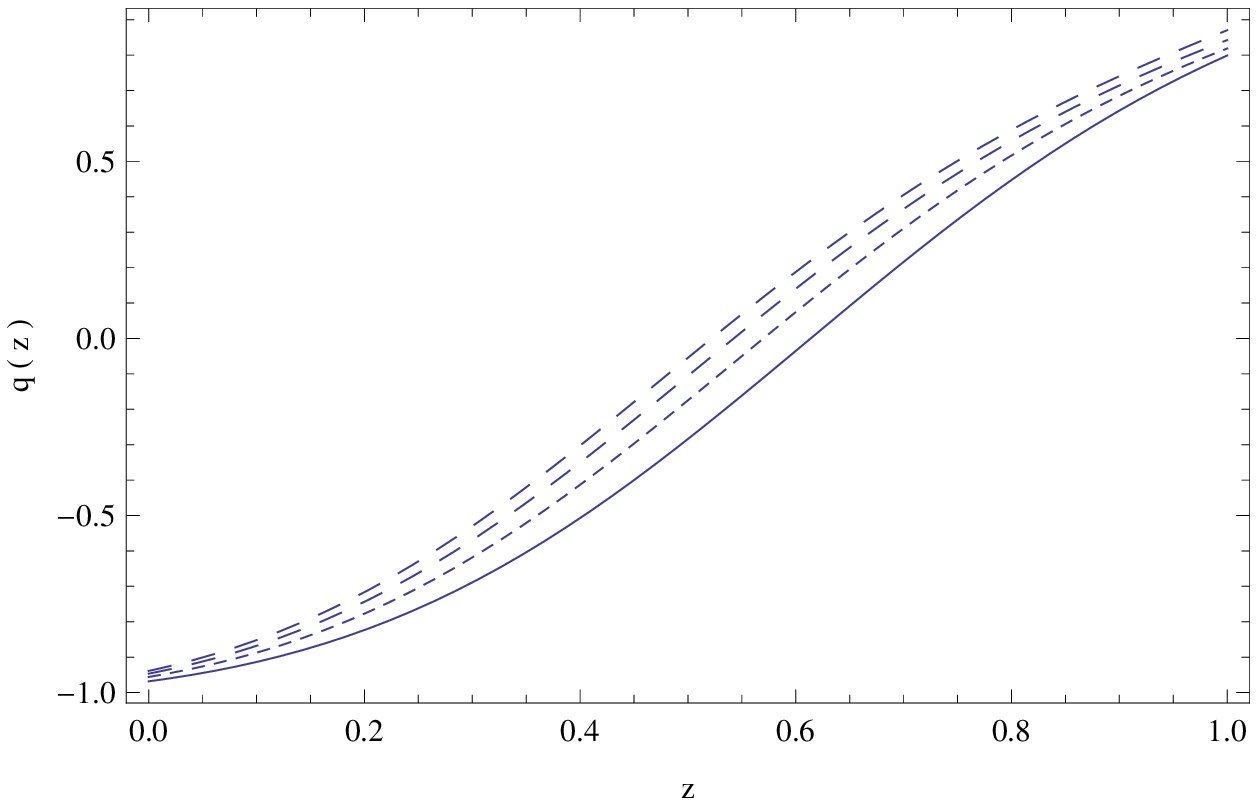}
   \hspace{0.5cm}
\includegraphics[width=8.15cm]{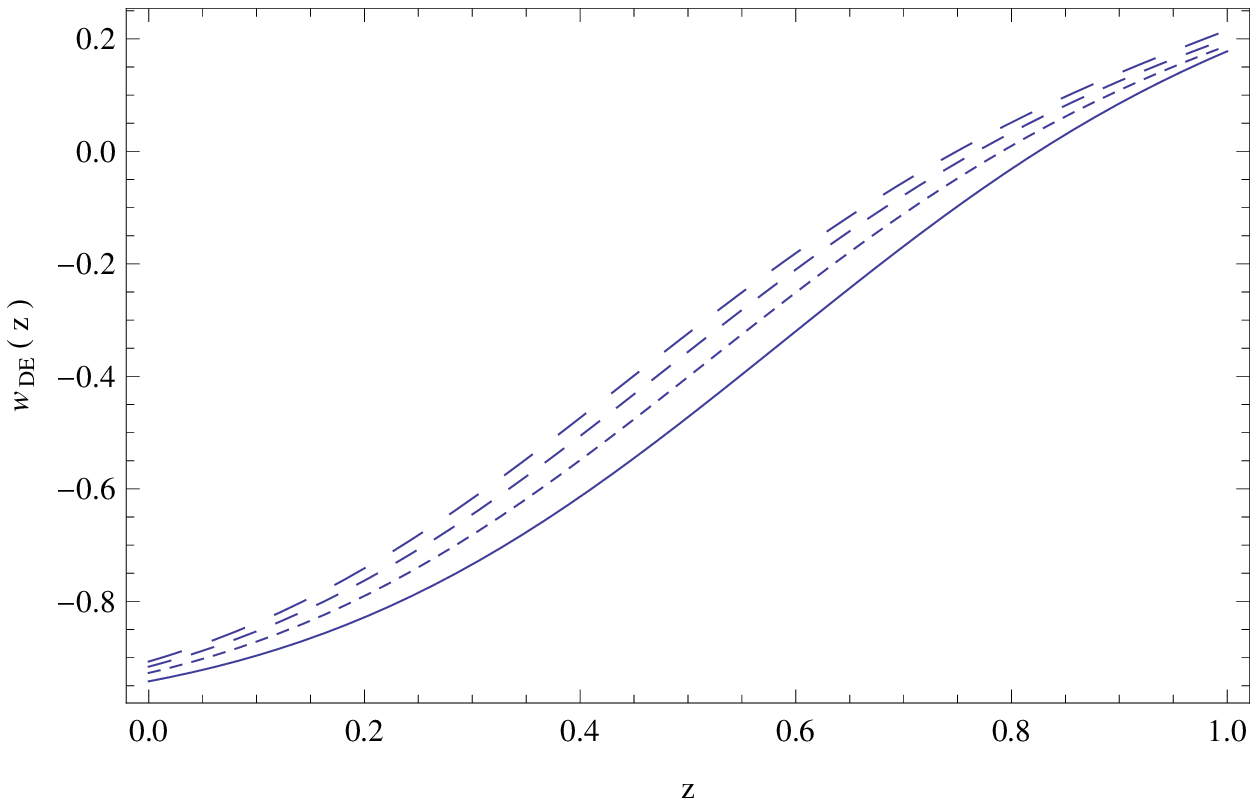}
\caption{
{\it{Evolution of the dimensionless Hubble function (top left), of the dimensionless field
(top right), of the deceleration parameter (bottom left), and of the dark-energy
equation-of-state parameter (bottom right), as a function of the redshift, for cosmology
with extended nonminimal derivative coupling in the case of the Higgs-like potential
(\ref{Higgspot2}) and for dust matter. The initial conditions have been chosen as
$h(0)=1$, $\Phi(0)=1$, and $\Pi (0)=0.1$, while the parameters of the exponential
potential have been fixed as $v_0=0.99$, $m=0.1$, $\lambda=0.1$. Concerning the coupling
parameters $\beta _0$ and $\zeta _0$, we choose: $\beta _0=1.90$ and $\zeta _0=1$ (solid
curve), $\beta _0=2.68$ and $\zeta _0=2$ (dotted curve), $\beta _0=3.29$ and $\zeta
_0=3$ (short dashed curve) and $\beta _0=3.79$ and $\zeta _0=4$ (dashed curve)
respectively.  }} }
\label{fig1}
\end{figure*}
\begin{eqnarray}
\frac{d\Pi }{dz} &=&-
\Big\{2\left( 1+z\right) h\left\{ h^{2}\left[ 5\beta
_{0}+\left( 5\beta _{0}^{2}-18\zeta _{0}\right) \Pi ^{2}-9\beta _{0}\zeta
_{0}\Pi ^{4}+6\zeta _{0}^{2}\Pi ^{6}\right] -5\beta _{0}\Pi ^{2}+3\zeta
_{0}\Pi ^{4}+15\right\} \Big\}^{-1}
\nonumber
\label{h2} \\
&&\times
\Bigg\{\left( 5\beta _{0}\Pi ^{2}-3\zeta _{0}\Pi ^{4}-15\right)
(-m^{2}\Phi +\Lambda \Phi ^{3})
-6h\Pi \left\{-2\Pi ^{2}\left[ 5\beta
_{0}+3\zeta _{0}\left( v_{0}-\frac{1}{2}m^{2}\Phi ^{2}+\frac{1}{4}\Lambda
\Phi ^{4}\right) \right] \right. \nonumber \\
&&\left. \ \ \ \ \,
+9\zeta _{0}\Pi ^{4}+5\beta _{0}\left( v_{0}-\frac{1}{2}m^{2}\Phi ^{2}+%
\frac{1}{4}\Lambda \Phi ^{4}\right) +15\right\}\Bigg\},
\end{eqnarray}
respectively.
Moreover, the dark-energy equation-of-state parameter (\ref{wDE}) reads as
\begin{eqnarray}
&&\!\!\!\!\!\!\!\!\!\!\!\!\!\!
w_{DE}=
\left\{\Pi
^{2}\left[ 1+h^{2}\left( \beta _{0}-\zeta _{0}\Pi ^{2}\right) \right]
+v\left( \Phi \right) \right\}^{-1}
\Bigg\{\Pi ^{2}-v\left( \Phi \right) -\frac{\beta _{0}}{9}\Pi \left\{
\left[ -2\left( 1+z\right) h\frac{dh}{dz}+3h^{2}\right] \Pi -4(1+z)h^{2}%
\frac{d\Pi }{dz}\right\}
  \nonumber \\
&&\ \ \ \ \ \ \ \ \ \ \ \ \ \ \ \ \ \ \ \ \ \ \ \ \ \ \ \ \ \ \ \ \ \ \ \ \ \ \ \ \ \ \ \
\ \ \ \ \ \ \ \ \ \
+\frac{\zeta _{0}}{15}\Pi ^{3}\left\{ \left[ -2(1+z)h%
\frac{dh}{dz}+3h^{2}\right] \Pi -8(1+z)h^{2}\frac{d\Pi }{dz}\right\} \Bigg\}\,.
\end{eqnarray}
Similarly to the previous cases, we can see that $w_{DE}$ can acquire values
of the phantom regime, which is an advantage of the scenario at hand.

In order to study the cosmological evolution of the dust Universe in the presence of the
Higgs-like potential we integrate the system of Eqs.~(\ref{h0})-(\ref{h2}) numerically. We
choose the potential parameters as $v_0=0.99$, $m=0.1$, $\lambda=0.1$, and for the initial
conditions we set $h(0)=1$, $\Phi(0)=1$, and $\Pi (0)=0.1$, and similarly to the
previous subsections we restrict our analysis at late times, namely at the redshift range
$0\leq z \leq 1$.

In Fig.~\ref{fig1} we depict the evolution of the Hubble function, of the scalar field,
of the deceleration parameter, and of the dark-energy equation-of-state parameter, in
terms of the redshift, for various values of the coupling parameters $\beta _0$ and
$\zeta_0$.

The Hubble function, presented in the top left graph, is a monotonically increasing
function of the redshift, and it becomes almost a constant in the redshift range $0\leq
z\leq 0.1$. The dimensionless scalar field $\Phi $, presented in the top right graph, is a
monotonically decreasing function of $z$. For the considered range of parameters, the
deceleration parameter $q$, presented in the bottom left graph, acquires a value
$q\approx 0.6$ at $z=1$, and it decreases monotonically with respect to the redshift,
crossing the $q=0$ line at redshifts of the order of $z\approx 0.5-0.6$, while tending
to the value $-1$ at present. Finally, the dark-energy equation-of-state parameter,
depicted in the bottom right graph, has a value $w_{DE}\approx 0.2$ at $z=1$, and it tends
to $-1$ at $z=0$, implying that the dark energy sector behaves like a cosmological
constant at present.

\subsubsection{Cosmological evolution of a radiation fluid}

\begin{figure*}[ht]
\centering
\includegraphics[width=8.15cm]{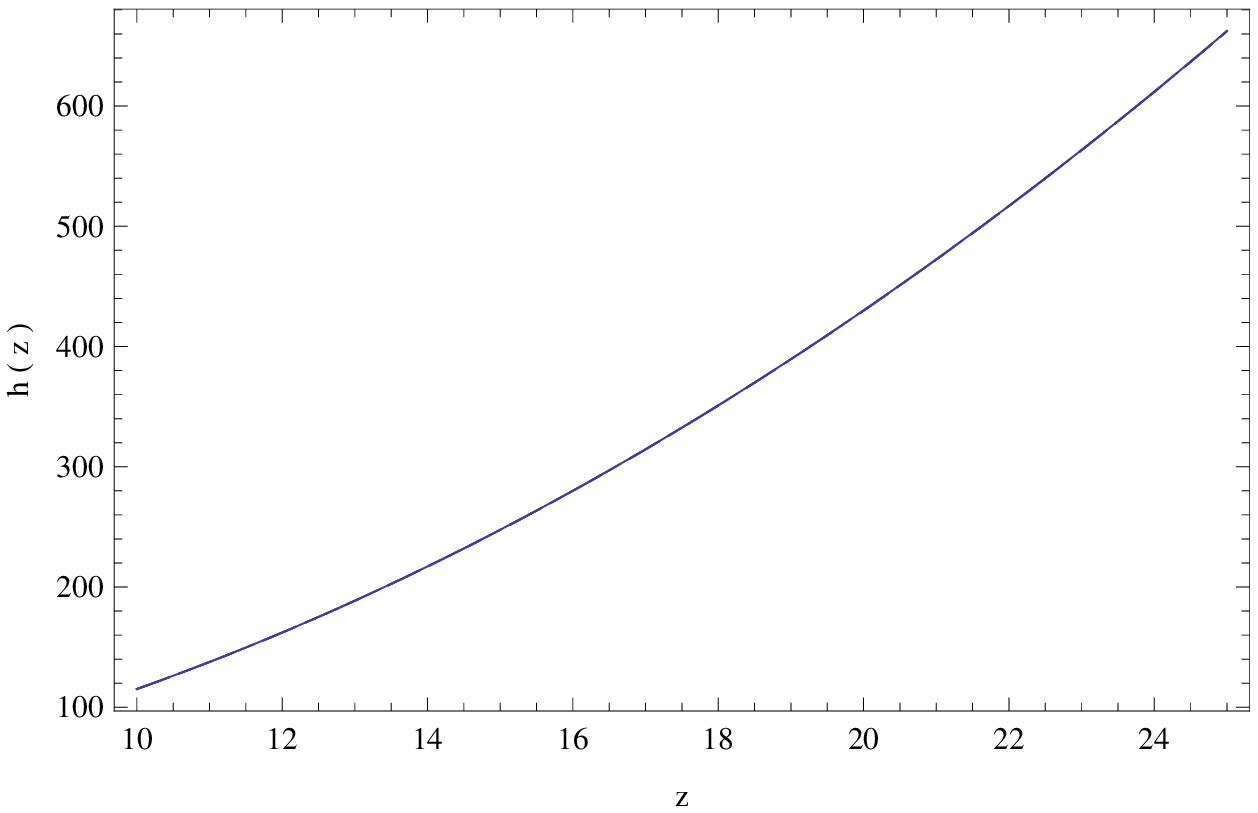} %
   \hspace{0.5cm}
\includegraphics[width=8.15cm]{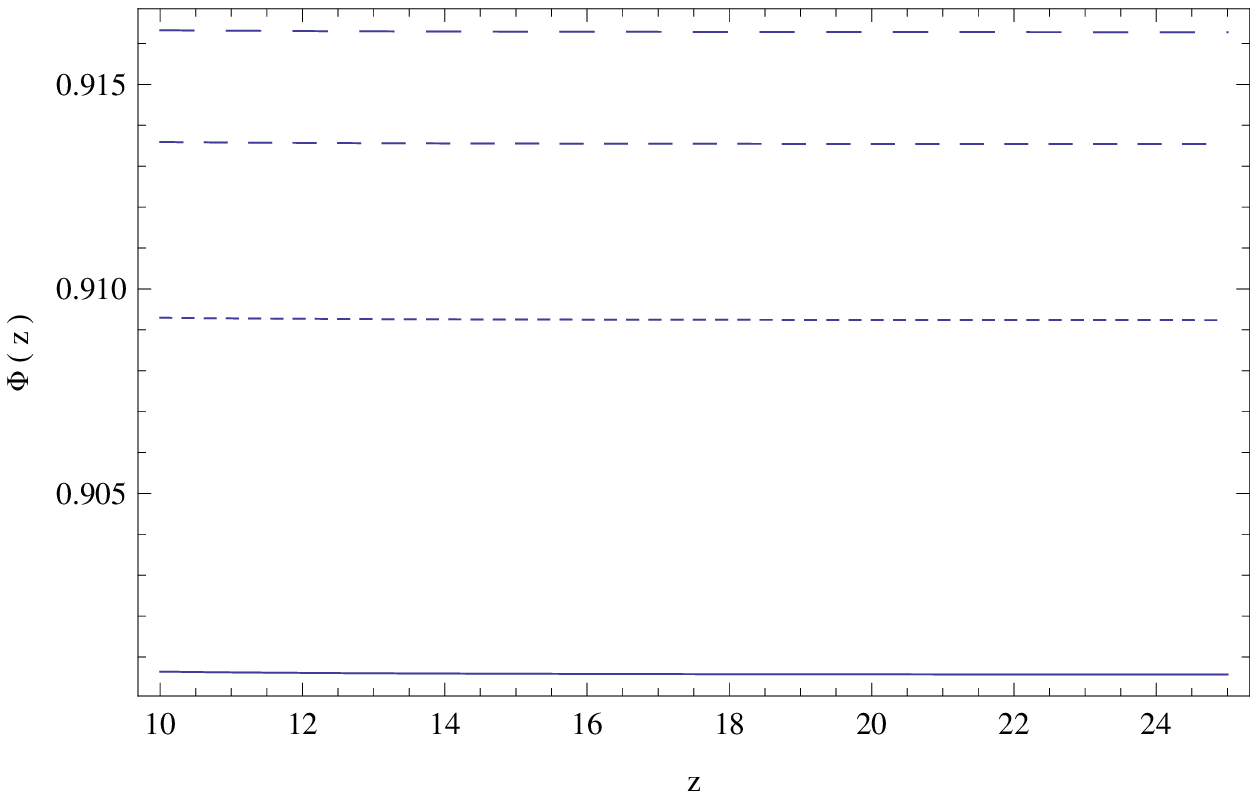} %
\includegraphics[width=8.15cm]{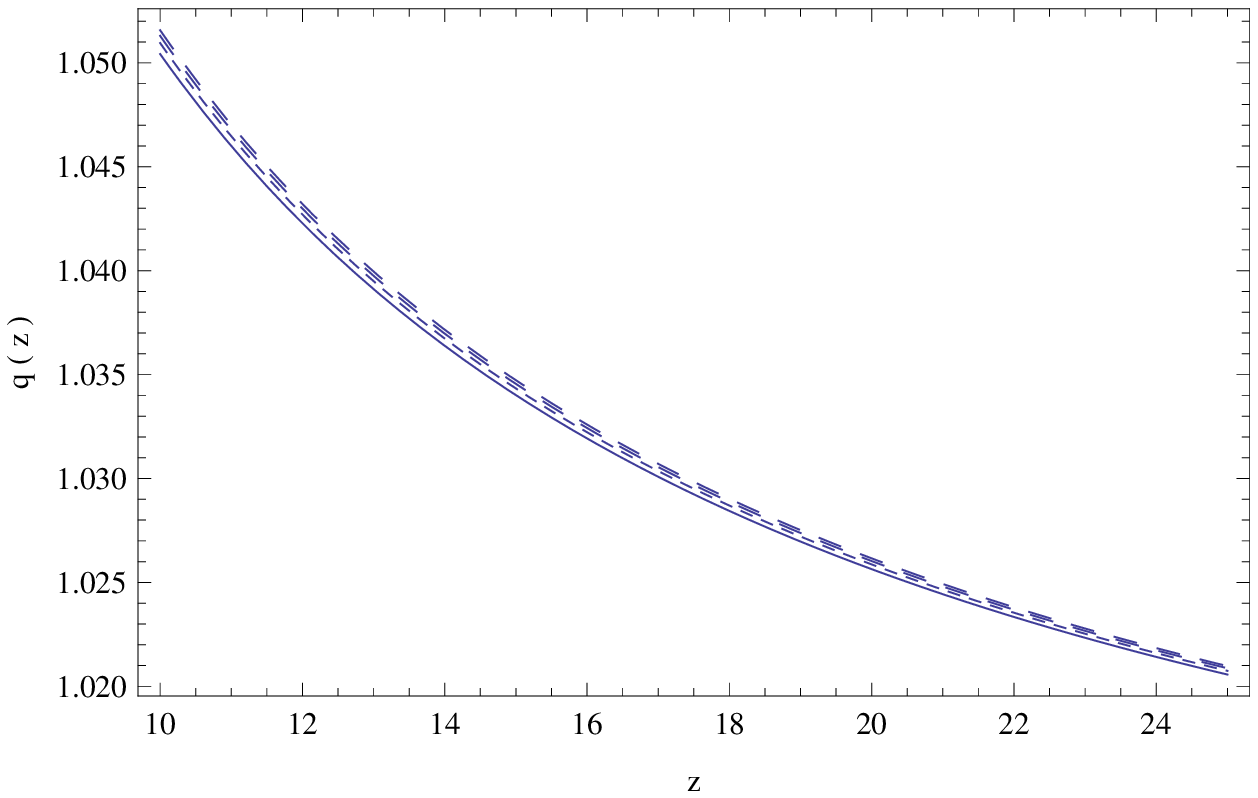}
   \hspace{0.5cm}
\includegraphics[width=8.15cm]{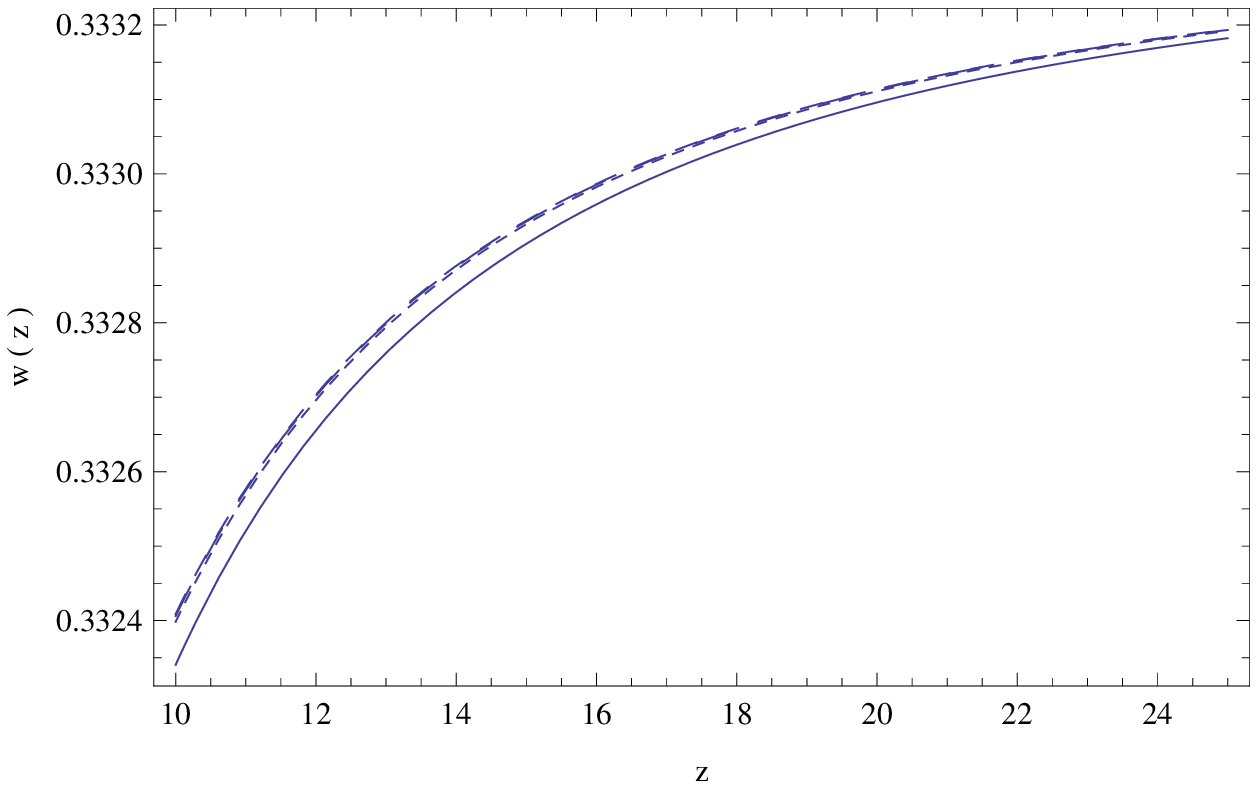}
\caption{
{\it{Evolution of the dimensionless Hubble function (top left), of the dimensionless
field (top right), of the deceleration parameter (bottom left), and of the total
equation-of-state parameter (bottom right), as a function of the redshift, for cosmology
with extended nonminimal derivative coupling in the case of the Higgs-like potential
(\ref{Higgspot2}) and for radiation fluid. The initial conditions have been chosen
as $h(0)=1$, $\Phi(0)=1$, and $\Pi (0)=0.1$, while the parameters of the exponential
potential have been fixed as $v_0=0.99$, $m=0.1$, $\lambda=0.1$. Concerning the coupling
parameters $\beta _0$ and $\zeta _0$, we choose: $\beta _0=1.897$ and $\zeta _0=1$ (solid
curve), $\beta _0=2.683$ and $\zeta _0=2$ (dotted curve), $\beta _0=3.286$ and $\zeta
_0=3$ (short dashed curve) and $\beta _0=3.794$ and $\zeta _0=4$ (dashed curve)
respectively.  }}}
\label{fig2}
\end{figure*}

For completeness, in this subsection we investigate the case where the matter fluid
corresponds to radiation, a case which is useful for early and intermediate stages
of the Universe evolution. In particular, we consider an equation of state of the form
$p_m=\rho_m/3$, and thus the dimensionless matter density and pressure scale with
respect to the redshift according to $r(z)=(1+z)^4$ and $P(z)=(1+z)^4/3$, respectively.
Moreover, we restrict our analysis at intermediate times, that is we focus on the
redshift range $10\leq z \leq 25$. In Fig.~\ref{fig2}, we depict the evolution of the
Hubble function, of the scalar field, of the deceleration parameter, and of the total
equation-of-state parameter of the Universe, in
terms of the redshift, for various values of the coupling parameters $\beta _0$ and
$\zeta_0$.

As we can see, the Hubble function, presented in the top left graph, is a monotonically
increasing function of the redshift. In the considered redshift range the scalar field,
shown in the top right graph, obtains constant values. The deceleration parameter,
plotted in the bottom left graph, is positive, with a slight increase from $q=1$ at $z=25$
to $q=1.05$ at $z=10$, indicating a decelerating behavior, as expected. Finally, the total
equation of state of the Universe, depicted in the bottom right graph, is positive, with
values of the order of $w\approx 0.33$, indicating a radiation-dominated expansion. We
mention that all quantities, apart from the scalar field, do not have a strong dependence
on the change of the numerical values of the  coupling parameters  $\beta _0$ and
$\zeta_0$, and hence the corresponding individual theories would not be easily
distinguishable.

\subsubsection{The unified picture of the evolution of the Universe in
theories with extended
nonminimal derivative coupling, in the presence of the Higgs potential}

\begin{figure*}[ht]
\centering
\includegraphics[width=8.15cm]{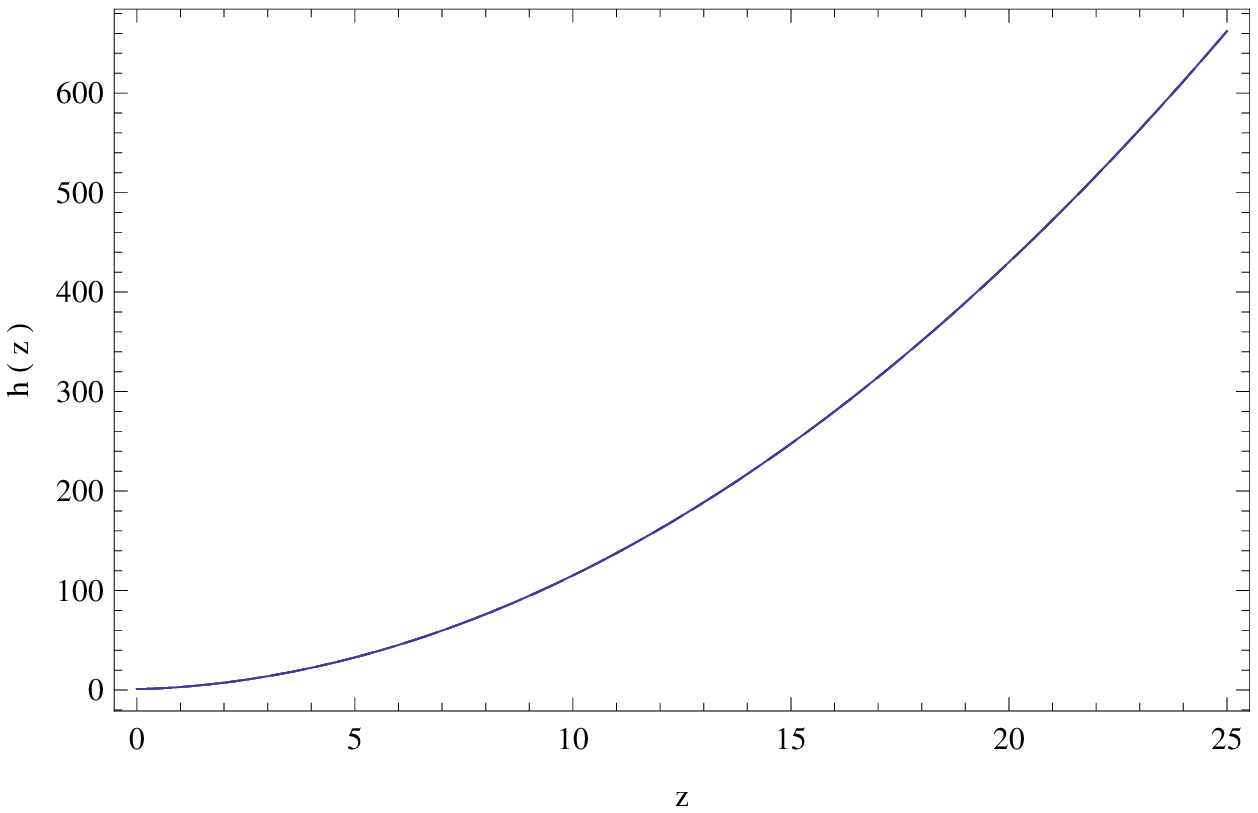} %
   \hspace{0.5cm}
\includegraphics[width=8.15cm]{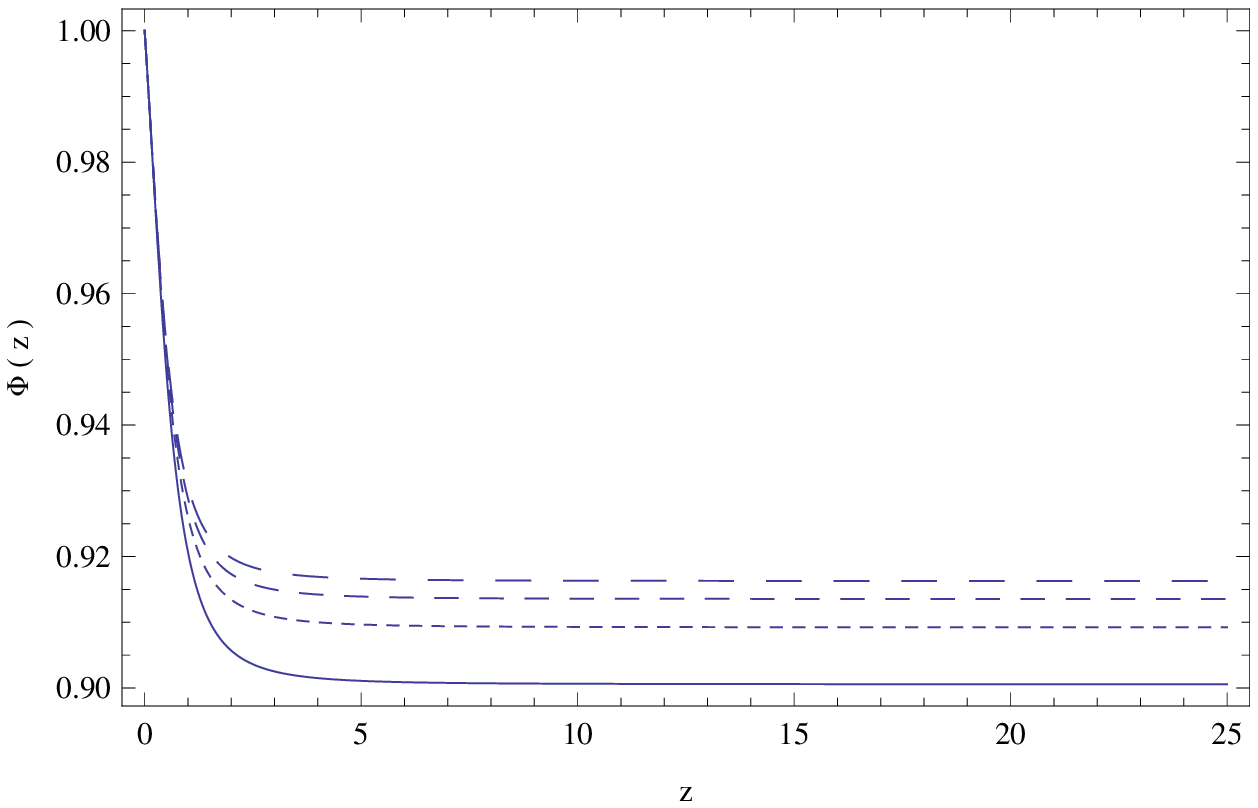} %
\includegraphics[width=8.15cm]{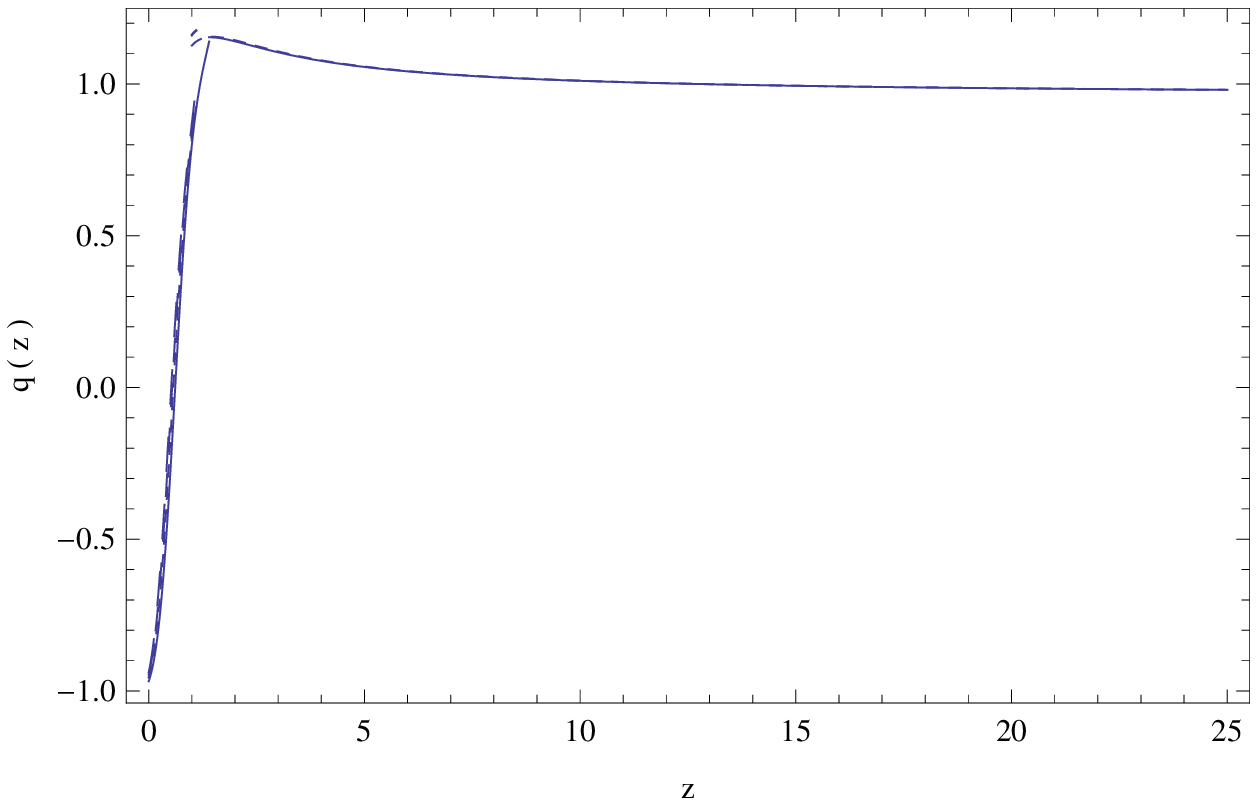}
   \hspace{0.5cm}
\includegraphics[width=8.15cm]{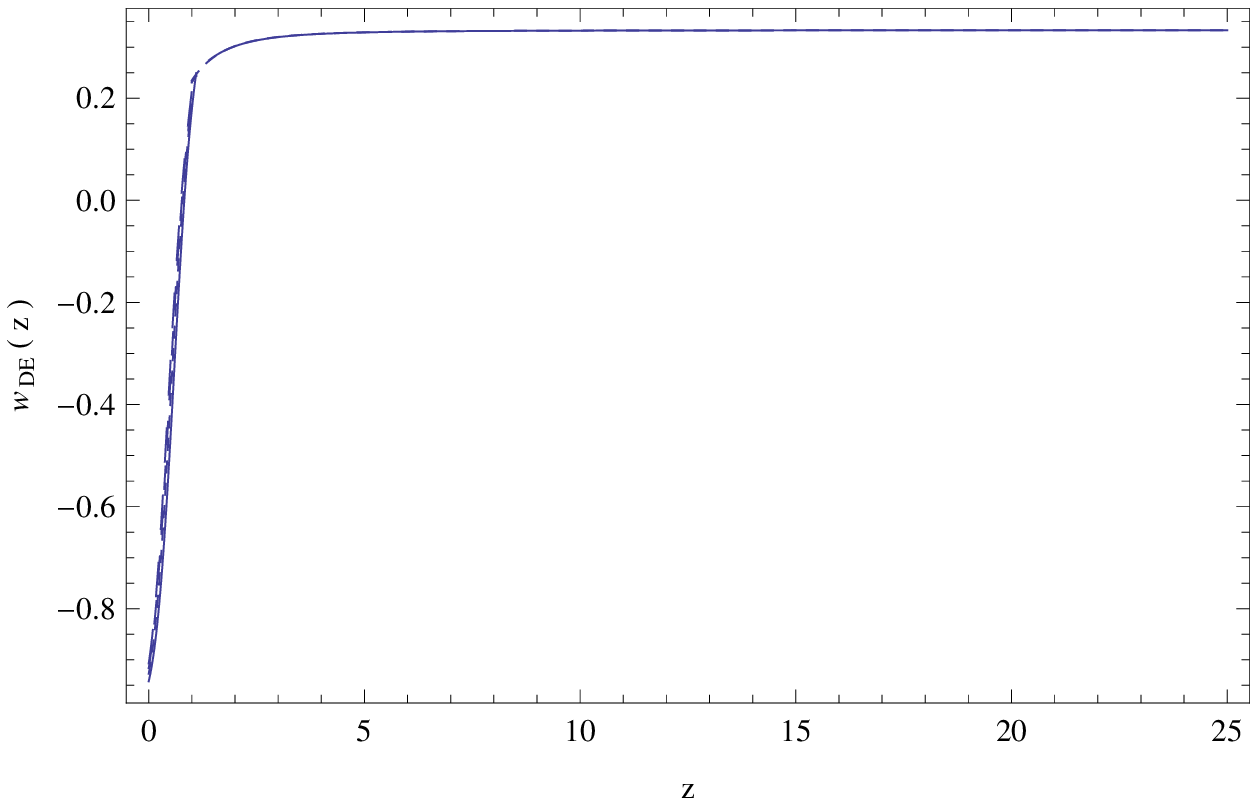}
\caption{
{\it{Evolution of the dimensionless Hubble function (top left), of the dimensionless
field (top right), of the deceleration parameter (bottom left), and of the total
equation-of-state parameter (bottom right), as a function of the redshift, for cosmology
with extended nonminimal derivative coupling in the case of the Higgs-like potential
(\ref{Higgspot2}), in the redshift range $z\in [0,25]$. The initial conditions have been chosen
as $h(0)=1$, $\Phi(0)=1$, and $\Pi (0)=0.1$, while the parameters of the exponential
potential have been fixed as $v_0=0.99$, $m=0.1$, $\lambda=0.1$. Concerning the coupling
parameters $\beta _0$ and $\zeta _0$, we choose: $\beta _0=1.897$ and $\zeta _0=1$ (solid
curve), $\beta _0=2.683$ and $\zeta _0=2$ (dotted curve), $\beta _0=3.286$ and $\zeta
_0=3$ (short dashed curve) and $\beta _0=3.794$ and $\zeta _0=4$ (dashed curve)
respectively.  }}}
\label{fig5}
\end{figure*}

Finally, to conclude the investigation of the  cosmological implications of theories with extended
nonminimal derivative coupling, in the presence of the Higgs potential, we present a unified
picture of the evolution of the Universe for the redshift
range $z \in \left(0,25\right)$. The variations of the Hubble function,
of the dimensionless scalar field, of the deceleration parameter, and of the
parameter of the dark energy equation of state are plotted in Figs.~\ref{fig5}, respectively. 

To obtain a unified picture of the evolution of the Universe in the presence of a gravitational theory with extended
nonminimal derivative coupling, and in the presence of a Higgs-type potential, we take for the redshift $z$  the range from 0
to 25. We further assume that in the range $z \in [5,25]$ the matter
content of the Universe can be described (at least approximately)
by a radiation type equation of state $p\approx \rho/3$. At the redshift
$z \approx 5$ the Universe enters in the matter dominated era, with
$p\approx 0$. In the present simplified model the transition from the
radiation dominated era to the matter dominated stage is
smooth, with all physical, geometrical and thermodynamical quantities
continue at the $z=5$ transition redshift. The Hubble
function and the scalar field $\Phi$, represented in the upper panels of Figs.~\ref{fig5} 
 are monotonically increasing and decreasing functions of the
redshift for the entire period. The evolution of the Hubble function is not significantly influenced by the variation of the model parameters. The scalar field $\Phi$ is approximately a constant in the redshift range $z\in (3,25]$, and its numerical values are strongly dependent on the model parameters. For redshifts $z<3$ the scalar field starts to increase, reaching its maximum value at $z=0$. In the range $z\in [0,2)$ the variation of the field is basically independent on the model parameters. The deceleration parameter, depicted in the bottom left panel of Figs.~\ref{fig5}, is approximately constant and positive in the redshift range $z\in (5,25]$, with numerical values of the order of $q\approx 1$. After the beginning of the matter dominate phase at $z=5$, the deceleration parameter increases, indicating a further deceleration of the Universe. But at $z\approx 2$, the Universe enters in an accelerating phase, and at $z=0$ the Universe experiences an exponential, de Sitter type expansion, with $q=-1$. The parameter $w_{DE}$ of the dark energy equation of state (bottom right panel of Figs.~\ref{fig5}) shows a similar dynamics. $w_{DE}$ is approximately constant in the reshift range $z\in (3,25]$, with $w_{DE}\approx 0.30$. At $z\approx 3$, $w_{DE}$ begins to decrease rapidly, and reaches the value $w_{DE}=-1$ at $z=0$, indicating that at this redshift the Universe is dominated by the effective dark energy generated by the extended nonminimal derivative coupling in the presence of a Higgs-type scalar field  potential. The cosmological evolutions of both $q$ and $w_{DE}$ are basically independent on the variation of the numerical values of the model parameters.  

\section{Conclusions}\label{sec5}

In this work we considered gravitational modifications that go beyond Horndeski, namely
we presented theories with extended nonminimal derivative coupling, in which the
coefficient functions depend not only on the scalar field but on its kinetic energy too.
Such theories prove to be ghost-free in a cosmological background, and hence it is
interesting to examine their cosmological implications. We first analyzed the cosmology
of these novel gravitational modifications at early times, neglecting the matter sector,
and we showed that a de Sitter inflation can be  realized even in the absence of a
potential term or of an explicit cosmological constant, and hence it is a pure result
of the extended gravitational couplings.

Additionally, we studied the behavior of these cosmological scenarios at late times,
where we obtained an effective dark energy sector arisen from the scalar field and its
extended couplings to gravity. We extracted various cosmological observables such as the
Hubble  function, the deceleration parameter, and the dark energy equation-of-state
parameter, and we numerically investigated their evolution at small redshifts, for three
choices of potentials, namely for the exponential, the power-law, and the Higgs one. As
we
showed, in all cases the Universe passes from deceleration to acceleration in the recent
cosmological past, while the effective dark-energy equation-of-state parameter tends to
the cosmological-constant value at present, in agreement with observations. Moreover, we
showed that the phantom regime can be accessible too, which is an advantage of the
scenarios since it is obtained despite the scalar field is canonical, i.e it
results purely from the novel, extended gravitational couplings.

The above features indicate that theories with extended nonminimal derivative could be a
good candidate for the description of early and late time Universe. Hence one could
proceed to more detailed analyses. In particular, one could use observational data from
Type Ia Supernovae (SNIa), Baryon Acoustic Oscillations (BAO), and Cosmic Microwave
Background (CMB) in order to constrain the coefficient functions, as well as the new
coupling parameters. Additionally, one could perform a complete dynamical analysis, in
order to by-pass the non-linearities of the equations, and extract the global behavior at
asymptotically late times. Moreover, one should analyze the perturbations in a thorough
way, in order to extract the values for inflation-related observables such as the
spectral index and the tensor-to-scalar ratio. Furthermore, the issue of the influence of the scalar degree of freedom in local gravity is an open problem, however it lies beyond the scope of the present work. It would be interesting to examine whether there are any issues related to the fifth force, as it has been done previously on the original Horndeski theory \cite{DeFelice:2011th,Koyama:2013paa}. Finally, it could be interesting to apply extended nonminimal derivative couplings to bi-scalar theories, such as those proposed recently in \cite{Ohashi:2015fma,Naruko:2015zze,Saridakis:2016ahq,Saridakis:2016mjd}. These investigations lie beyond the scope of the present work, and are left for future
projects.

\begin{acknowledgments}
We thank Nelson Nunes for helpful comments and suggestions. FSNL acknowledges financial
support of the Funda\c{c}\~{a}o para a Ci\^{e}ncia e Tecnologia through an Investigador
FCT Research contract, with reference IF/00859/2012, funded by FCT/MCTES (Portugal). This
article is based upon work from COST Action ``Cosmology and Astrophysics Network
for Theoretical Advances and Training Actions'', supported by COST (European Cooperation
in Science and Technology).
\end{acknowledgments}



\end{document}